\documentclass[%
 aip,
 amsmath,amssymb,
 reprint,%
]{revtex4-1}

\usepackage{textcomp}
\usepackage{soul}       % provides \st (strike-through)
\setstcolor{red}        % make the strike-through line red
\usepackage{graphicx}% Include figure files
\usepackage{dcolumn}% Align table columns on decimal point
\usepackage{bm}% bold math
\usepackage{xcolor}
\usepackage[utf8]{inputenc}
\usepackage[T1]{fontenc}
\usepackage{mathptmx}
\usepackage{etoolbox}
\makeatletter
\def\@email#1#2{%
 \endgroup
 \patchcmd{\titleblock@produce}
  {\frontmatter@RRAPformat}
  {\frontmatter@RRAPformat{\produce@RRAP{*#1\href{mailto:#2}{#2}}}\frontmatter@RRAPformat}
  {}{}
}%
\makeatother

\makeatletter
\def\@email#1#2{%
 \endgroup
 \patchcmd{\titleblock@produce}
  {\frontmatter@RRAPformat}
  {\frontmatter@RRAPformat{\produce@RRAP{*#1\href{mailto:#2}{#2}}}\frontmatter@RRAPformat}
  {}{}
}%
\makeatother

\makeatletter
\renewcommand\section{\@startsection {section}{1}{\z@}%
  {-1.5ex \@plus -1ex \@minus -.2ex}%
  {0.5ex \@plus.2ex}% 
  {\normalfont\normalsize\bfseries\raggedright}}% Left-aligned section
\renewcommand\subsection{\@startsection {subsection}{2}{\z@}%
  {-1.25ex\@plus -1ex \@minus -.2ex}%
  {0.3ex \@plus .2ex}%
  {\normalfont\normalsize\bfseries\raggedright}}% Left-aligned subsection
\makeatother

\makeatletter
\patchcmd{\@sect}{\MakeUppercase}{\relax}{}{}
\patchcmd{\@ssect}{\MakeUppercase}{\relax}{}{}
\makeatother

\begin{document}
\preprint{AIP/123-QED}

%%%%%%%%%%%%% AUTHORS %%%%%%%%%%%%%%%%%%%%%

\author{Giulia Meucci\textsuperscript{*}}
\affiliation{Department of Energy Conversion and Storage, Technical University of Denmark, 2800 Kgs.\ Lyngby, Denmark}  \email{giuliam@dtu.dk}

\author{Pinelopi O. Konstantinopoulou}
\affiliation{Department of Energy Conversion and Storage, Technical University of Denmark, 2800 Kgs.\ Lyngby, Denmark}

\author{Thies Jansen}
\affiliation{Department of Energy Conversion and Storage, Technical University of Denmark, 2800 Kgs.\ Lyngby, Denmark}

\author{Gunjan Nagda}
\affiliation{Department of Energy Conversion and Storage, Technical University of Denmark, 2800 Kgs.\ Lyngby, Denmark}

\author{Damon J. Carrad}
\affiliation{Department of Energy Conversion and Storage, Technical University of Denmark, 2800 Kgs.\ Lyngby, Denmark}

\author{Emiliano Di Gennaro}
\affiliation{Dipartimento di Fisica "Ettore Pancini", Università degli Studi di Napoli “Federico II”,  Complesso Universitario di Monte S. Angelo, Via Cinthia, 80126 Naples, Italy}

\author{Yu Chen}
\affiliation{CNR-SPIN, c/o Complesso di Monte S. Angelo, via Cinthia, 80126  Naples, Italy}

\author{Nicola Manca}
\affiliation{CNR-SPIN, C.so F. M. Perrone 24, 16152 Genova, Italy}

\author{Nicolas Bergeal}
\affiliation{Laboratoire de Physique et d’Etude des Matériaux, ESPCI Paris, Université PSL, CNRS, Sorbonne Université, Paris, France}

\author{Manuel Bibes}
\affiliation{Laboratoire Albert Fert, CNRS, Thales, Université Paris-Saclay, 91767 Palaiseau, France}

\author{Alexei Kalaboukhov}
\affiliation{Department of Microtechnology and Nanoscience (MC2), Chalmers University of Technology, 41296 Gothenburg, Sweden}

\author{Marco Salluzzo}
\affiliation{CNR-SPIN, c/o Complesso di Monte S. Angelo, via Cinthia, 80126  Naples, Italy}

\author{Roberta Citro}
\affiliation{Dipartimento di Fisica “E.R. Caianiello,” Università di Salerno, 84084 Fisciano, Italy} \affiliation{CNR-SPIN, c/o Universita di Salerno, 84084 Fisciano, Italy}

\author{Felix Trier}
\affiliation{Department of Energy Conversion and Storage, Technical University of Denmark, 2800 Kgs.\ Lyngby, Denmark}

\author{Nini Pryds}
\affiliation{Department of Energy Conversion and Storage, Technical University of Denmark, 2800 Kgs.\ Lyngby, Denmark}

\author{Fabio Miletto Granozio}
\affiliation{CNR-SPIN, c/o Complesso di Monte S. Angelo, via Cinthia, 80126  Naples, Italy}

\author{Alessia Sambri}
\affiliation{CNR-SPIN, c/o Complesso di Monte S. Angelo, via Cinthia, 80126  Naples, Italy}

\author{Thomas S. Jespersen}
\affiliation{Department of Energy Conversion and Storage, Technical University of Denmark, 2800 Kgs.\ Lyngby, Denmark}\affiliation{Center For Quantum Devices, Niels Bohr Institute, University of Copenhagen, 2100 Copenhagen, Denmark}

\date{\today}

%%%%%%%%%%%%% TITLE %%%%%%%%%%%%%%%%%%%%%
\title{Intrinsic Negative-U Centers in Freestanding LaAlO$_3$/SrTiO$_3$ Micro-membranes}

\begin{abstract}
The LaAlO$_3$/SrTiO$_3$ (LAO/STO) interface hosts a rich range of electronic phenomena, including unconventional electron pairing that in quantum dots gives rise to a negative effective charging energy $U$.  Here, we show freestanding LAO/STO micro-membranes  naturally hosting negative-$U$ centers, where lateral confinement arises intrinsically, rather than from engineered nanostructures. These centers coexist with gate-tunable superconductivity and can remain stable upon thermal cycling from millikelvin temperatures to room temperature. Transport is 
in excellent agreement with calculations based on a negative-$U$ Anderson model, and electrostatic simulations indicate characteristic center sizes of 20–80 nm.  Our findings suggest that negative-$U$ centers may arise from the intrinsic interfacial inhomogeneity typical of LAO/STO, and should therefore be considered a general feature of the LAO/STO interface. This could have important consequences for the microwave response of interfacial superconducting devices.

\end{abstract}

\maketitle

%%%%%%%%%%%%% Figure 1 %%%%%%%%%%%%%%
\begin{figure*}[hbt]
\centering
\includegraphics[width = \linewidth]{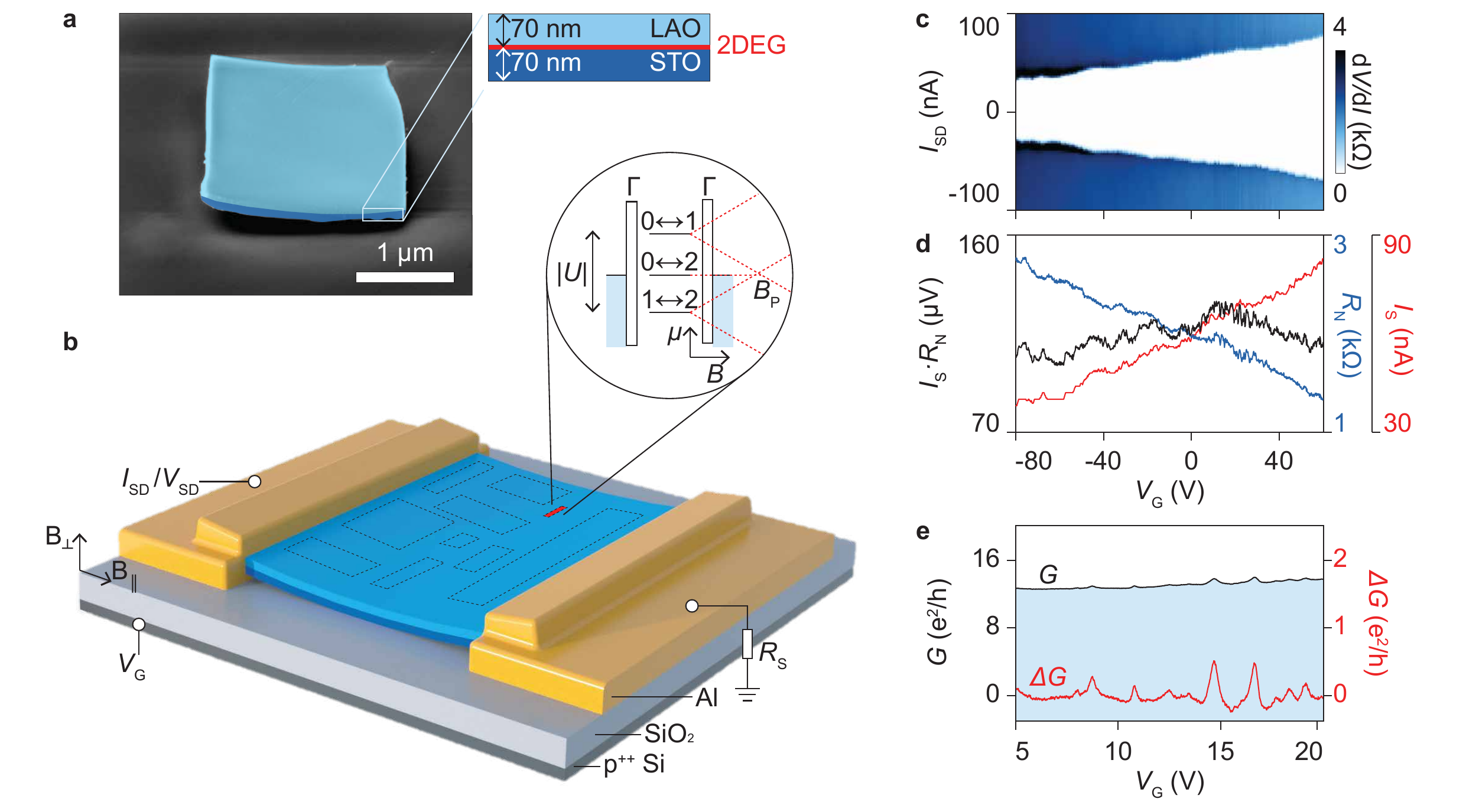}
\caption{
\textbf{a} False color scanning electron micrograph of a micro-membrane on SiO$_2$.
\textbf{b} Device schematic  together with  a  representation of the array of conductive puddles at the LAO/STO interface (not to scale). A nanometer-sized puddle (red) may represent the intrinsic origin of the observed negative-$U$ center.  Inset: electrochemical potentials of the leads and  transitions in a negative-$U$ center (black) and their evolution with magnetic field (red). 
\textbf{c} Differential resistance of Dev\#1 as a function of $V_{\text{G}}$ and $I_{\text{SD}}$, with $B_{\parallel} = 25\ \mathrm{mT}$, to suppress Al-contact superconductivity ($T_{\text{c}}^{\text{Al}}\approx$1.2 K,  $B_{\text{c}}^{\text{Al}}\approx$10 mT). 
\textbf{d} Gate dependence of switching current $I_{\text{s}}$,  normal-state resistance $R_{\text{N}}$ and their product  $I_{\text{s}} \cdot R_\text{N}$.
\textbf{e} Normal-state conductance at $B_{\perp} = 0.5\ \mathrm{T}$ vs $V_\text{G}$ (black trace). The red trace shows the residual $\Delta G$ after background subtraction.
}
 % \hfill
\end{figure*}

%%%%%%%%%%%%% Figure 2 %%%%%%%%%%%%%%
\begin{figure*}[hbt]
\centering
\includegraphics[width = \linewidth]{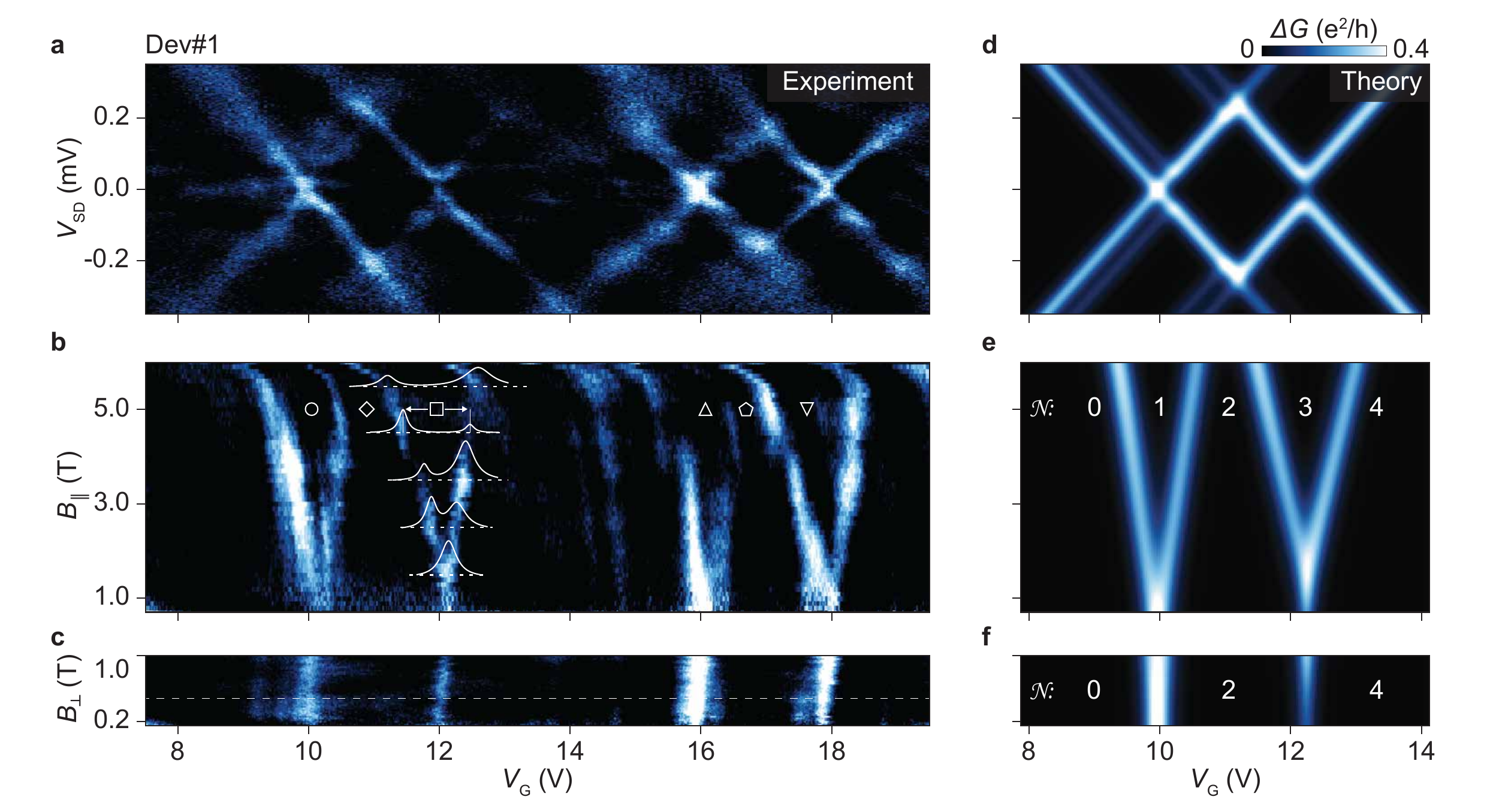}
\caption{
    \textbf{a} Measured residual conductance, $\Delta G$, of Dev\#1 as a function of $V_\text{G}$ and $V_\text{SD}$, at $B_{\perp} = 0.5\ \mathrm{T}$ (white dashed line in panel c). 
    \textbf{b}  Measured zero-bias $\Delta G$ versus $V_\text{G}$ and $B_{\parallel}$.   The white curves show Lorentzian fits of a selected peak/peak pair at  $B_{\parallel}=1.5,\ 2.5,\ 3.5,\ 4.5,\ 5.5\ \mathrm{T}$. The symbols correspond to those in Fig.~3.
    \textbf{c} Measured zero-bias $\Delta G$ versus $V_\text{G}$ and $B_{\perp}$.   
    Panels b and c are shifted along  $V_\text{G}$ by $\approx1.5\ \mathrm{V}$ and $\approx1.3\ \mathrm{V}$ respectively, to align the low-field peak center with panel a; the shift arises from device drift during measurements. 
     \textbf{d} Simulated  conductance versus $V_\text{G}$ and $V_\text{SD}$.
    \textbf{e}  Simulated zero-bias  conductance  versus $V_\text{G}$ and $B_{\parallel}$.   
    \textbf{f} Simulated zero-bias  conductance  versus $V_\text{G}$ and $B_{\perp}$.  The numbers on panel e and f indicate the ground-state charge occupancy $\mathcal{N}$.
}
\end{figure*}

%%%%%%%%%%%%% Figure 3 %%%%%%%%%%%%%%
\begin{figure}[tb]
\centering
    \includegraphics[width = \linewidth]{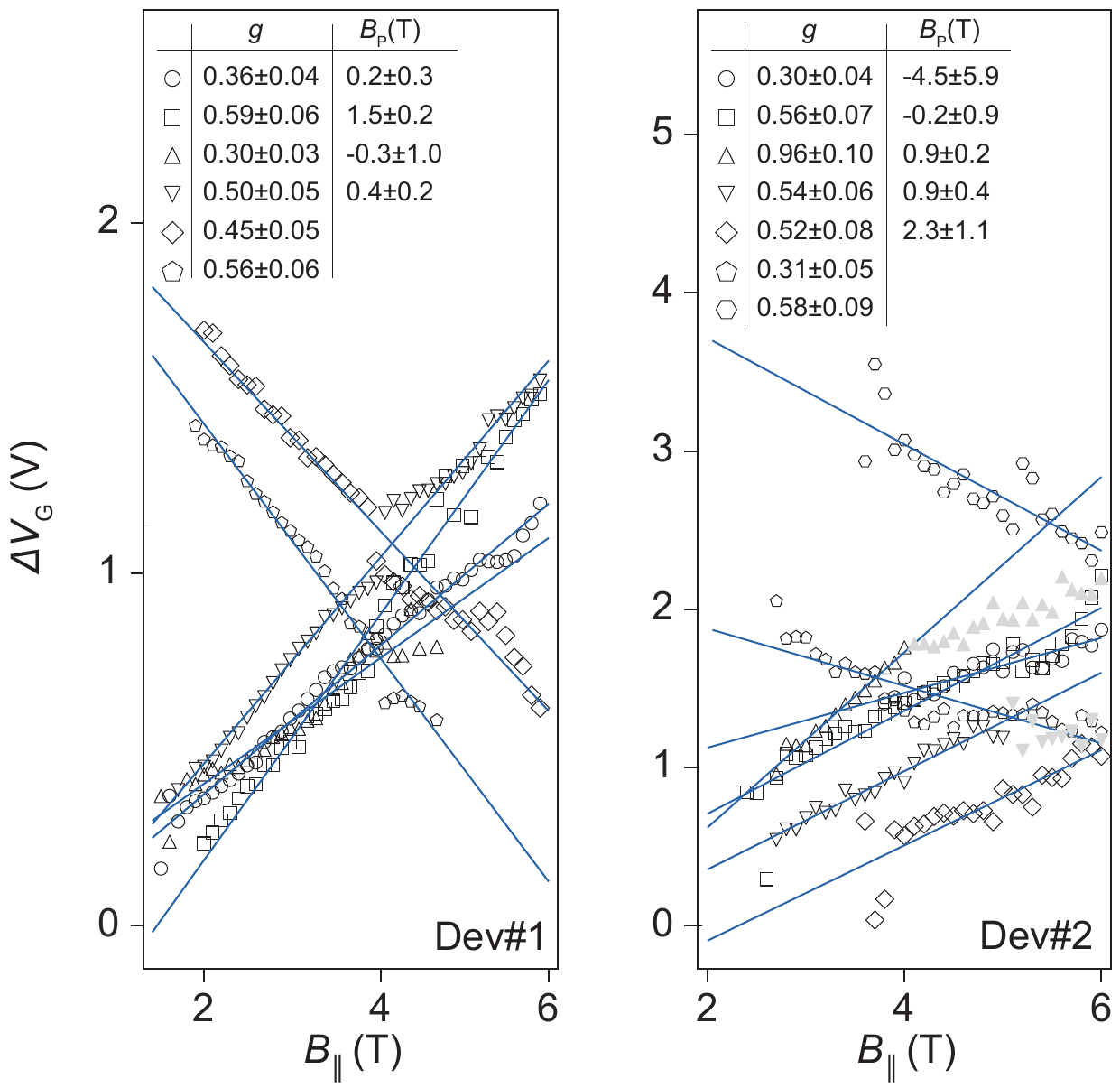 }
    \caption{ $B_{\parallel}$ dependence of the gate voltage separation between zero-bias conductance peaks in Dev\#1 and Dev\#2. The symbols correspond to those in Fig.~2b and S3.  The pairing field, $B_{\text{P}}$, and the Landé $g$-factor,  $g$, are extracted from linear fits (solid lines). For two peak pairs of Dev\#2,  data points at high fields (filled grey markers) are excluded from the fit. Note that in some cases the very low slope of $\Delta V_\text{G}(B)$ renders $B_\text{P}$ ill-defined.} 
 % \hfill
\end{figure}

%%%%%%%%%%%%% Figure 4 %%%%%%%%%%%%%%
\begin{figure}[b]
\centering
    \includegraphics[width = \linewidth]{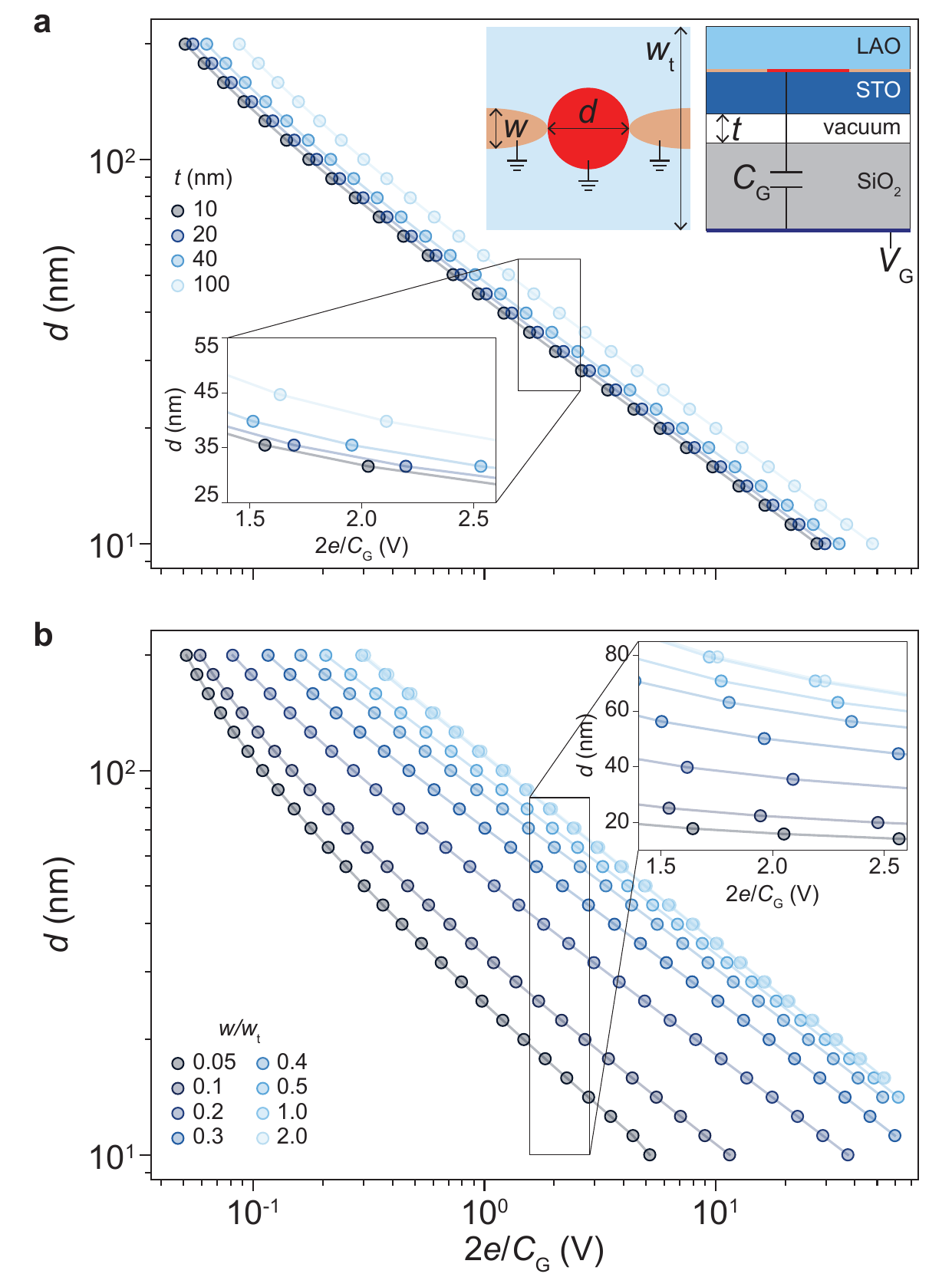 }
    \caption{ \textbf{a} Estimated center diameter as a function of gate capacitance $C_\text{G}$ for a contact aspect ratio $w/w_\text{t} = 0.2$ and various vacuum thicknesses $t$. The inset shows a schematic of the simulated geometry (top and side views). 
    \textbf{b} Estimated center diameter as a function of $C_\text{G}$ for a fixed vacuum thickness $t=50\ \mathrm{nm}$ and varying contact aspect ratios $w/w_\text{t}$.
    }
 % \hfill
\end{figure}

%%%%%%%%%%%%% Figure 5 %%%%%%%%%%%%%%
\begin{figure*}[t]
\centering
    \includegraphics[width = \linewidth]{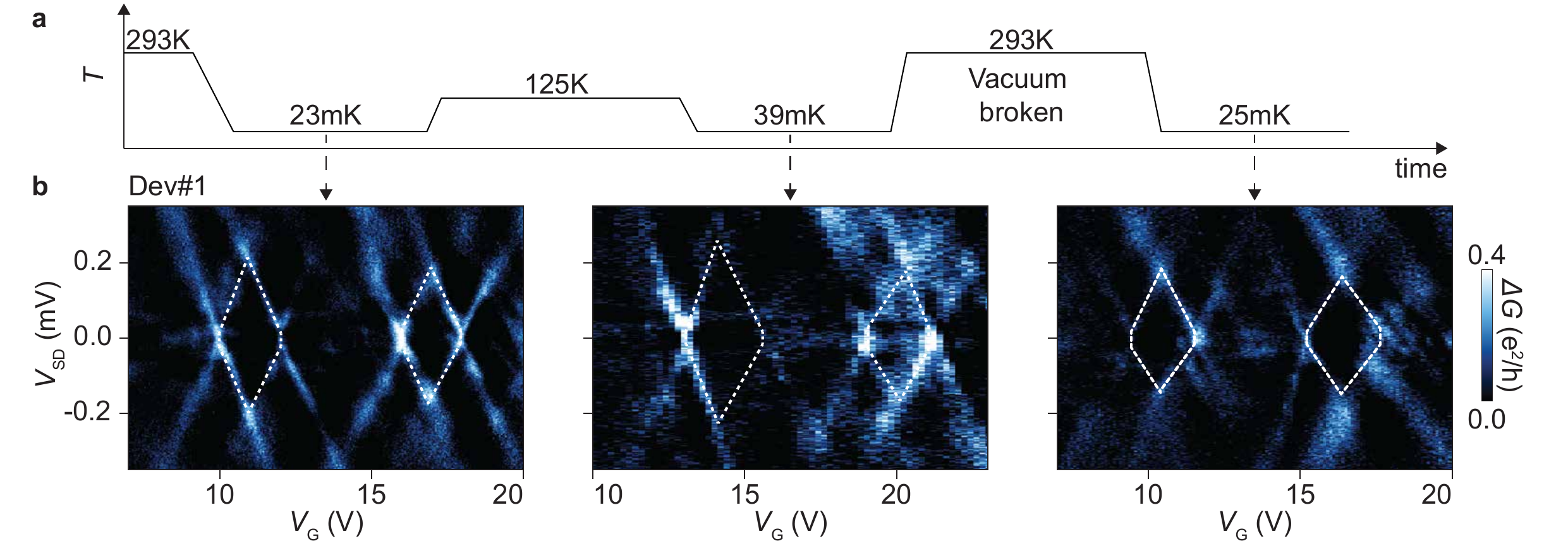}
    \caption{\textbf{a} Thermal cycling sequence applied to Dev\#1.
    \textbf{b} Residual conductance of Dev\#1 after the first cool-down, after thermal cycling to 125 K, and after thermal cycling to room temperature with exposure to air,  at $B_{\perp} = 0.5\ \mathrm{T}$. White dashed lines are guides to the eye.  }
\end{figure*}

Interfaces between complex oxides provide a rich platform for exploring correlated electron phenomena. A notable example is the two-dimensional electron gas (2DEG) at the interface between LaAlO$_3$ (LAO) and SrTiO$_3$ (STO), which exhibits a remarkable combination of properties, including high mobility \cite{ohtomo2004high}, strong Rashba spin–orbit coupling \cite{caviglia2010tunable}, ferromagnetic ordering \cite{ christensen2019strain}, negative compressibility \cite{li2011very} and gate-controlled superconductivity \cite{thiel2006tunable, caviglia2008electric, erlandsen2022two, reyren2007superconducting}. Its low temperature phase diagram resembles that of high-temperature superconductors, with a dome-shaped superconducting phase and  a softly gapped state above the superconducting critical temperature $T_\text{c}$ \cite{caviglia2008electric, richter2013interface}.  Electron pairing has also been directly observed  at temperatures above $T_\text{c}$ and  magnetic fields exceeding the superconducting critical field  $B_\text{c}$  \cite{cheng2015electron,  prawiroatmodjo2017transport, bjorlig2020g, annadi2018quantized, briggeman2020pascal}. In this regime  electron-electron interactions in quantum dots are modeled by a  negative effective charging energy $U$ in the Anderson model \cite{anderson1975model}, in contrast to the positive Coulomb repulsion  of conventional semiconductor quantum dots. \\
\indent Previous negative-$U$ quantum dot experiments rely on extrinsic confinement imposed by electrostatic gating\cite{prawiroatmodjo2017transport, bjorlig2020g} or atomic-force microscopy (AFM)-based sketching\cite{cheng2015electron}. However, numerous studies provide strong evidence that even in pristine epitaxial LAO/STO the electronic landscape is intrinsically inhomogeneous, and low-temperature transport is mediated by submicrometric highly conductive “puddles” embedded in a less conductive background 
\cite{ kalisky2013locally, honig2013local,   caprara2012intrinsic, caprara2013multiband,  bovenzi2015possible,  scopigno2016phase}. In the superconducting phase, the puddles host local superconductivity and couple via the Josephson effect \cite{biscaras2013multiple,  hurand2019josephson, prawiroatmodjo2016evidence, manca2019}.  Several mechanisms have been proposed to account for this intrinsic inhomogeneity, including electronic instability associated with negative compressibility and the non-rigidity of the interfacial confinement potential \cite{scopigno2016phase, bovenzi2015possible}, charge segregation mediated by strong Rashba spin–orbit coupling \cite{caprara2012intrinsic},  interaction of the interfacial electrons with oxygen vacancies \cite{ piyanzina2019oxygen}, and influence of the ferroelastic domain-wall network of STO on the  2DEG \cite{kalisky2013locally}.\\
\indent Despite a wealth of experiments on millimeter-scale LAO/STO 2DEGs, insight into the microscopic origins of emergent behavior is hampered by difficulties in fabricating devices at smaller scales \cite{schneider2006microlithography, stornaiuolo2012plane, minhas2017temperature, kalaboukhov2017homogeneous, d2021nanopatterning, singh2022gate}. Therefore, the  majority of transport studies have focused on as-grown unpatterned  heterostructures or macroscopic Hall-bar structures, probing the average properties of the inhomogeneous interface. Insight into  microscopic mechanisms is, however, crucial both for a fundamental understanding of the interface and for the development of STO-based micro-scale functional electronics. \\
\indent Freestanding oxide membranes \cite{lu2016synthesis} have opened a new path toward the 
integration of oxides with semiconductors \cite{lu2019freestanding} and flexible electronics \cite{bakaul2017high} with enhanced control over curvature and strain \cite{kum2019epitaxial}. Additionally, when scaled down to the micron-level\cite{sambri2020self, erlandsen2022two}, freestanding micro-membranes offer a promising platform  for realization of
micron-scale devices to probe the intrinsic inhomogeneity. \\
\indent Here, we study electronic devices fabricated from $\sim1  \times2$ \textmu m$^2$ freestanding LAO/STO (thickness: 70 nm / 70 nm) micro-membranes realized by spontaneous spalling\cite{bedell2013layer} and transferred onto a silicon oxide / silicon substrate (Fig.~1a).  Two devices were investigated in detail: Dev\#1 with the LAO layer facing the substrate and Dev\#2 with the opposite orientation (Supplementary Section S1). Growth details are presented in Ref.\ \citenum{sambri2020self}, and fabrication details are reported in Methods.  Magneto-transport measurements were performed in a dilution refrigerator at  temperature  $T_{\text{base}}\approx$ 20 mK, unless stated otherwise (Methods and Supplementary Section S2). Figure 1b shows a device schematic with magnetic field orientations: $B_{\parallel}$ parallel to the membrane plane and current flow, and $B_{\perp}$ perpendicular to the membrane. \\
\indent Although no extrinsic electrostatic confinement was implemented the as-cooled  sub-kelvin electronic properties show direct signatures of transport through intrinsic negative-$U$ centers, superimposed on a background of  gate-tunable superconductivity.  This indicates that negative-$U$ centers can emerge without engineered lateral confinement, suggesting an intrinsic origin related to the electronic inhomogeneity of the interfacial 2DEG. The results are well reproduced by rate-equation computations based on a negative-$U$ Anderson model. From the characteristic energies of the devices and finite element electrostatic simulations we estimate an effective lateral confinement size of $20-80\ \mathrm{nm}$. Finally, we find that the negative-$U$ confinement can be robust upon thermal cycling from millikelvin temperatures to room temperature, and we discuss the confinement origin  and its connection to the inhomogeneous electronic landscape at the LAO/STO interface.
\subsection*{\large{Gate-tunable superconductivity}}
\indent Figure 1c shows the differential resistance of Dev\#1 as a function of  gate voltage, $V_{\text{G}}$, and bias current, $I_{\text{SD}}$, revealing  a superconducting ground state as previously analyzed in Ref.\ \cite{erlandsen2022two}. The switching current, $I_{\text{s}}$, increases  with $V_\text{G}$ (Fig.~1d) and the corresponding 
current density, $J_c\approx15-45$ nA/\textmu m, is comparable to the highest values reported for high-quality epitaxial bulk samples \cite{hurand2019josephson}. The normal-state resistance, $R_\text{N}$, measured at $B_{\perp}=1\ \mathrm{T}$, decreases with increasing $V_\text{G}$, such that the product $I_{\text{s}} \cdot R_\text{N}\approx114 \pm7$ \textmu V remains constant. These observations are  consistent with a 2DEG described by a Josephson array as discussed in Refs.\ \citenum{biscaras2013multiple, prawiroatmodjo2016evidence, hurand2019josephson} (Fig.~1d).  This is further consistent with the gradual transition from  the normal to the zero-resistance state as a function of temperature (Supplementary Section S3). The gate dependence of $T_\text{c}$ and $B_\text{c}$ follows that of Ref.~\citenum{erlandsen2022two} and is presented in Supplementary Section S3. Over the range $-80~\text{V} \le V_\text{G} \le 60~\text{V}$, $T_\text{c}$ increases from 216 to 226 mK ($\pm 1\ \mathrm{mK}$), while $B_{\parallel}^\text{c}=815 \pm 15\ \mathrm{mT}$ and $B_{\perp}^\text{c}=150 \pm 10\ \mathrm{mT}$ remain constant within experimental error. The anisotropy of the critical field is consistent with the two-dimensional nature of transport, and we  estimate a coherence length $\xi \approx 47\ \mathrm{nm}$ and an upper bound of the superconducting thickness $d < 30\ \mathrm{nm}$ \cite{reyren2009anisotropy, ben2009anisotropic, erlandsen2022two} (Supplementary Section S3). The weak $V_\text{G}$-dependence of $T_\text{c}$, indicates that the superconducting gap is essentially gate-independent in the measured range.  Altogether, the extracted parameters agree with previous reports on bulk samples \cite{reyren2007superconducting, caviglia2008electric, ben2010tuning, prawiroatmodjo2016evidence} and freestanding micro-membranes \cite{erlandsen2022two}, and are consistent with a two-dimensional Josephson array of superconducting puddles, with $V_\text{G}$ mainly controlling the Josephson coupling between them \cite{hurand2019josephson}.  
\subsection*{\large{Evidence of confinement}}
\indent In addition to the overall decreasing trend, $R_\text{N}(V_\text{G})$ exhibits reproducible modulations. These are emphasized in Fig.~1e showing the conductance, $G=1/R_\text{N}$, measured at $B_{\perp} = 0.5 \ \mathrm{T}$  in a narrower $V_\text{G}$-range. Distinct conductance peaks are observed on the smoothly increasing background which is associated with increasing carrier density of the 2DEG. To enhance the modulations we subtract a polynomial background and Fig.~1e shows the residual conductance, $\Delta G$ (Supplementary Section S4).\\
\indent Figure 2a shows  $\Delta G$ as a function of $V_\text{G}$ and bias voltage $V_\text{SD}$. A  diamond pattern of high conductance is observed, resembling the features expected for transport through a  quantum dot in Coulomb blockade. This observation implies transport through discrete charge states of a nanoscale region with total capacitance $C$, giving a charging energy $E_\text{c}\approx e^2/C \gg k_{B}T$, where $e$ is the electron charge and $k_{B}$ is the Boltzmann constant. Further, the region must be coupled to the source and drain reservoirs via tunnel barriers with coupling strengths $\Gamma\ll E_\text{c}$ (inset of Fig.~1b). Within each conductance diamond, the total charge on the nanoscale region is fixed and can be written as  $Q=(\mathcal{N_\text{0}+N})e$, where $\mathcal{N_\text{0}}e$ is the charge at the lowest $V_\text{G}$ in Fig.~2a and  $\mathcal{N}=0,1,2, \dots$. Charge variation occurs only at the diamond boundaries, where transport becomes allowed. This finding is notable because, unlike previously reported quantum dots in LAO/STO, which were defined by electrostatic gates \cite{maniv2016tunneling, prawiroatmodjo2017transport, thierschmann2018transport, bjorlig2020g} or by AFM sketching \cite{cen2009oxide, cheng2015electron}, here there is no extrinsic confinement. Moreover, the device is operated at high electron densities, well away from depletion, as evidenced by the relatively high $T_\text{c}$ in the superconducting regime \cite{richter2013interface}.  A similar pattern is observed for Dev\#2 as presented in Supplementary Section S5. \\
\indent From the diamond $V_\text{SD}$, we extract an addition energy, $E_{\text{add}}\approx 190 \pm20$ \textmu eV for Dev\#1 and between 100 and 420 \textmu eV, depending on the charge state,  for Dev\#2.  The lever arm, $\alpha=\text{d}V_\text{SD}/\text{d}V_\text{G}\approx (1.0 \pm 0.1)  \cdot 10^{-4}$ for both devices, quantifies the ratio of capacitive coupling between source/drain and gate. The uncertainties are related to broadening of the diamond boundaries. These estimates assume that the applied $V_{\text{SD}}$ drops only across the quantum dot; any series resistance would reduce the voltage drop across the center.  The extracted values should therefore be seen as upper limits.  
\subsection*{\large{Negative-$U$ behavior}}
\indent Figures 2b-c show the magnetic field dependence of the zero-bias conductance, for $B_{\parallel}$ and $B_{\perp}$ respectively. Data at $B \ll B_{\text{c}}$ are excluded as the conductance peaks are masked by the superconducting state (Supplementary Section S4).  As $B_{\parallel}$ increases, the peaks remain initially field-insensitive before bifurcation at a pairing field $B_{\text{P}}$, after which they split linearly with field (Fig.~2b). For $B_{\perp}$ the peaks remain unaffected up to the 1 T limit of our setup (Fig.~2c). While Coulomb blockade is widely reported and well understood in semiconductor systems and metallic nanoparticles, the observed doubling of Coulomb blockade resonances in field is unusual. It indicates that the zero-field charge ground state consists of even parity  states ($\mathcal{N} = 0, 2, 4, \dots$), with odd states becoming accessible only at $B > B_{\text{P}}$ (inset of Fig.~1b). This behavior is consistent with transport through a negative-$U$ center, featuring effective attractive interaction and a paired ground state persisting until $B_{\text{P}}$, as previously observed in patterned LAO/STO devices \cite{cheng2015electron, prawiroatmodjo2017transport, bjorlig2020g}. The results of Fig.~2 show that negative-$U$ centers can also occur as  intrinsic features of the unpatterned LAO/STO interface. \\
\indent For each peak pair, the peak separation $\Delta V_\text{G}(B)$ was estimated by fitting $G(V_\text{G})$ to a double-Lorentzian line-shape (Fig.~3, see Supplementary Section S6 for details). The Landé $g$-factor was obtained from the slope of linear fits to $\Delta V_\text{G}(B)$ as $g =\frac{1}{\mu_{\text{B}}}\frac{\text{d}}{\text{d}B} (e \alpha \Delta V_{\text{G}})$, where $\mu_\text{B}$ is the Bohr magneton, while $B_{\text{P}}$ was determined by extrapolating $\Delta V_\text{G}(B)$ to zero. From the extracted values, listed in Fig.~3, we estimate~$U = g \mu_\text{B} B_\text{P}$ to fall within the range from $-20$ to $-40$ \textmu eV for both devices.  These values are consistent with previously reported negative-$U$ interactions in zero-\cite{cheng2015electron, prawiroatmodjo2017transport} and one- \cite{annadi2018quantized, briggeman2020pascal} dimensional systems with engineered confinement, though slightly smaller in magnitude.
\subsection*{\large{Transport simulations}}
\indent To model the experimental data, we used the extracted $U$, $g$, and $\alpha$ to construct a two-orbital Anderson Hamiltonian for the negative-$U$ center tunnel-coupled to the leads. Transport was calculated by solving the Pauli master equation and the results for two resonances of Dev\#1 are shown in Fig. 2d–f as a function of $V_\text{G}$, $V_\text{SD}$, $B_{\parallel}$ and $B_{\perp}$ (Supplementary Section S7). The simulations closely reproduce the experimental data, demonstrating that the intrinsic centers observed in the freestanding micro-membranes can be well described by a negative-$U$ Anderson model, similarly to extrinsically defined quantum dots in LAO/STO. 
\subsection*{\large{Electrostatic simulations of the center size}}
\indent To gain further insight into the nature of the intrinsic confinement defining the negative-$U$ centers, we performed electrostatic simulations to estimate the  their size and conducted thermal cycling experiments to investigate the robustness of the confinement. The effective size was estimated from the gate potential $\Delta V_\text{G}=2e/C_\text{G}$ required to charge the center with two electrons via the gate. We performed electrostatic simulations of $C_\text{G}$ modeling the system as a stack of 70 nm LAO, 70 nm STO, a possible vacuum separation of thickness $t$ between the membrane and the substrate, and 500 nm SiO$_2$ (see schematics in Fig.~4a). The numerical simulations include the electric field dependence of the dielectric permittivity of STO \cite{davidovikj2017quantum}. Varying the gate voltage or reversing the LAO/STO order does not significantly affect the results, as the SiO$_2$ layer dominates $C_\text{G}$ (Supplementary Section S8).  In all the reported simulations we set $V_\text{G}=10\ \mathrm{V}$.  The negative-$U$ center is modeled as a grounded circular disk of diameter $d$ at the LAO/STO interface, and the source and drain electrodes as grounded half ellipses with semi-axis $w$ and different ratios relative to the membrane width $w_\text{t}$. The accumulated charge on the disk, $Q$, and the resulting capacitance, $C_\text{G}=Q/V_\text{G}$, were calculated for various combinations of  $d$, $t$ and $w/w_\text{t}$. Figure 4a shows the corresponding two-electron charging voltage $\Delta V_\text{G}$  as a function of $d$ for several $t$ values at fixed $w/w_\text{t}=0.2$, while Fig.~4b  shows $\Delta V_\text{G} (d)$ for various $w/w_\text{t}$ at $t=50\ \mathrm{nm}$.  Matching the simulated values to the experimental  $\Delta V_\text{G}\approx2\ \mathrm{V}$ yields center diameters between 20 and 80 nm, depending on the parameters selected. 
\subsection*{\large{Thermal stability}}
\indent Finally we consider the reproducibility of the results upon thermal cycling. Remarkably, unlike the tetragonal domain structure in thin-film LAO/STO\cite{kalisky2013locally, honig2013local},  
the ($V_\text{SD}$, $V_\text{G}$) stability diagram of Dev\#1 remains largely unaffected after thermal cycling $T_\text{base} \rightarrow125\text{ K}\rightarrow T_\text{base}$ and even after an additional cycle $T_\text{base}\rightarrow 300\text{ K}\rightarrow T_\text{base}$, including exposure to ambient conditions at room temperature (Fig.~5). On the other hand, in Dev\#2, the conductance diamonds disappear after cycling the temperature to 125 K (Supplementary Section S5).\\
\indent The thermal stability of at least one device shows that confinement is not induced by extrinsic effects such as surface contamination or mobile impurities, which are known to induce quantum dots in semiconductor nano-channels close to depletion \cite{scott1989conductance}. Not only do the micro-membrane devices operate far from depletion, but  such extrinsic potentials would also randomize upon thermal cycling. Structural defects, such as oxygen vacancies or interstitials, induce potential modulations that are thermally stable below room temperature. However, this mechanism seems  unlikely here, since the device operates at large electron densities, and, thus, the defect potential would need to locally deplete the 2DEG while simultaneously leaving a conductive, electrically isolated region within the depleted area. We are not aware of common defects with such dual action.   Non-uniform strain fields, arising from the membrane roughness, could generate local potential variations. However, electrostatic simulations indicate that the typical roughness on the STO side, which is the less smooth of the two surfaces, cannot generate potential modulations strong enough to confine charge (Supplementary Section S9). 
\subsection*{\large{Discussion}}
\indent The presented observations can be naturally interpreted in the context of the intrinsic electronic inhomogeneity of the LAO/STO interface, commonly described by the puddle picture of transport. In this framework, charge carriers could localize on weakly coupled nanometer-sized puddles, acting as confinement centers. Josephson-array modeling of the superconducting phase has inferred puddle dimensions  on the order of 100 nm \cite{biscaras2013multiple, prawiroatmodjo2016evidence, hurand2019josephson}, consistent with the sizes extracted here for negative-$U$ centers. In addition, the superconducting properties of freestanding micro‑membranes closely match those of thin‑film LAO/STO\cite{erlandsen2022two}. Furthermore, both puddle formation and negative-$U$ behavior are intrinsic to the interface.\\
\indent The development of freestanding oxide membranes is central to current research and technology, and the results presented here show that in these systems negative-$U$ centers can appear intrinsically. Similar behavior may therefore be expected in membranes prepared by other methods \cite{lu2016synthesis, eom2021electronically}. It is also plausible that these centers originate from interface inhomogeneity and, thus, are not specific to freestanding micro-membranes, but rather a general feature of the LAO/STO interface. The reduced spatial averaging in the micro-membrane devices leaves fewer puddles contributing to transport compared to bulk samples, making individual localized centers a significant contribution to the total conductance. This is consistent with  reports on thin-film LAO/STO submicron‑scale devices which exhibit a qualitative change in behavior for lateral dimensions below $\approx1$ \textmu m, observations which were attributed  to the intrinsic interface inhomogeneity \cite{aurino2016retention, minhas2017temperature}.  \\
\indent An open question concerns the origin of interfacial inhomogeneity and its possible link to ferroelastic domains in STO. Although domain sizes typically exceed those of the puddles\cite{kalisky2013locally, honig2013local}, domain walls may still act as pinning sites for them. In this context, the high strain in freestanding membranes could modify domain dynamics during thermal cycling, with strain differences between devices potentially accounting for their different thermal‑cycling behavior.  Alternatively, strain could stabilize monodomain states\cite{sambri2020self}, pointing to inhomogeneity mechanisms unrelated to structural domains\cite{caprara2012intrinsic, caprara2013multiband, bovenzi2015possible, scopigno2016phase}. Clarifying domain structure in micro-membranes must therefore be the subject of future studies. \\
\indent In summary, we have reported evidence of weakly gate-tunable superconductivity and negative-$U$ centers in freestanding LAO/STO micro-membranes on SiO$_2$/Si. Unlike previous observations in engineered nanostructures, the negative-$U$ behavior here emerges intrinsically, without imposed lateral confinement. Transport simulations based on a negative-$U$ Anderson model reproduce the data with $U \approx -20$ to $-40$ \textmu eV, and electrostatic modeling suggests center sizes of 20–80 nm. One center remains stable after thermal cycling from millikelvin temperatures to room temperature. These results demonstrate that negative-$U$ centers can form spontaneously at the LAO/STO interface, consistent with an intrinsically inhomogeneous electronic landscape. The nanoscale confinement required for their formation may therefore be a general property of the interface, connected to the same intrinsic inhomogeneity  thought to underlie superconductivity in large-scale devices modeled as random Josephson arrays. Our findings may thus offer new opportunities for a deeper understanding of the electronic landscape at the LAO/STO interface, which in turn may help uncover the origin and consequences of both the electronic inhomogeneity and the negative‑$U$ character of this system.  The gate-tunable ground state of the LAO/STO offers a unique potential for realizing  reconfigurable superconducting electronics, and while the membrane geometry eliminates the bulk volume of STO which imposes large dielectric losses \cite{manca2019, singh2018competition}, our findings show that the  2DEG itself hosts intrinsic two level systems which can affect the loss tangent of superconducting devices even far from depletion.  \\

\bibliography{ref}
\vspace{0.5cm}
\subsection*{\large{Methods}}

\textbf{Device fabrication}\\
For device fabrication, micro-membranes were transferred  from the growth substrate onto a p$^{++}$ Si substrate  capped with 500 nm  of SiO\textsubscript{2}, serving as gate-dielectric. Electrical contacts to the interfacial 2DEG were defined by electron-beam lithography and argon ion Kaufmann milling at 50° with respect to the membrane plane,  followed by in-situ Ti/Al (4 nm / 200 nm) evaporation at 70° from two antiparallel in-plane orientations, and 200 nm Al evaporated at 90°.  \\

\noindent\textbf{Magneto-transport measurements}\\
Magneto-transport measurements were performed in a dilution refrigerator with base temperature  $T_{\text{base}}\approx$ 20 mK.  The series resistance from cryostat filters (6–11 k$\Omega$) and the contacts (<500 $\Omega$) was measured from the baseline in the superconducting regime and subtracted (Supplementary Section S2). The cryostat was equipped with a vector magnet capable of applying magnetic fields up to 6 T in the direction parallel to the membrane plane and current flow ($B_{\parallel}$), and up to 1 T perpendicular to the membrane ($B_{\perp}$) .   \\
\indent \textit{Measurements in Fig.~1c:} A current bias was applied to one electrode, and the voltage was measured at the same electrode, while the other electrode was grounded. The DC current bias, $I_\text{SD}$, was supplied using a voltage source and a 10~M$\Omega$ resistor. An AC excitation of 4 nA was superimposed using the voltage output of a lock-in amplifier and a 1~G$\Omega$ resistor. The AC excitation frequency was 283 Hz. Both AC and DC voltages were measured using a voltage amplifier with a gain of 100.  From the AC voltage, the differential resistance was computed as $dV/dI$. \\
\indent \textit{Measurements in Fig.~1e, Fig.~2, and Fig.~5:} A voltage bias was applied to one electrode, and the current was measured at the opposite electrode. The DC voltage bias, $V_\text{SD}$,  was supplied using a voltage source through a $10^3$ divider. Due to instrumental drift, the zero of the DC bias presented an offset, drifting over time; therefore, the plots were corrected for vertical offsets between 20 and 80 \textmu eV. An AC excitation of 30 \textmu V was superimposed using the voltage output of a lock-in amplifier and a $10^5$ divider. The AC excitation frequency was 283 Hz.  AC and DC currents were measured using a current-to-voltage converter with a gain of $10^6$.  From the AC current, the differential conductance was computed as $G=dI/dV$. \\

\noindent\textbf{Transport simulations}\\
Details of the transport simulations are described in Supplementary Section S7. \\

\noindent \textbf{Electrostatic simulations} \\
The electrostatic simulations were performed using COMSOL. Details are provided in the main text and in Supplementary sections S8, S9. The COMSOL files used for the simulations are available in the data repository.\\

\subsection*{\large{Data availability} }
The data that support the findings of this study are openly available at 10.11583/DTU.30893570.\\

\subsection*{\large{Acknowledgments}}
This work was supported by the European Research Council through the European Union’s Horizon 2020 research and innovation program (Grant No. 866158) and through the Horizon Europe EIC Pathfinder program Project IQARO (Grant No. 101115190). We also acknowledge support from Ministero dell’Istruzione, dell’Università e della
Ricerca for the PRIN PNRR 2022 project FOXES (Prot. P2022TCT72), for the PRIN 2022 project OMEGA (Prot. 2022TCJP8K) and for the PNRR Project PE0000023-NQSTI. F. T. acknowledges support from Villum Fonden via research grant 37338 (SANSIT). N.P. acknowledges funding from the ERC Advanced (NEXUS, grant no. 101054572) and the support from Villum Fonden via research grant VIL73726. \\

\subsection*{\large{Competing interests}}
The authors have no conflicts to disclose.\\

\end{document}

\author{Giulia Meucci}
\affiliation{Department of Energy Conversion and Storage, Technical University of Denmark, 2800 Kgs.\ Lyngby, Denmark}

\author{Pinelopi O. Konstantinopoulou}
\affiliation{Department of Energy Conversion and Storage, Technical University of Denmark, 2800 Kgs.\ Lyngby, Denmark}

\author{Thies Jansen}
\affiliation{Department of Energy Conversion and Storage, Technical University of Denmark, 2800 Kgs.\ Lyngby, Denmark}

\author{Gunjan Nagda}
\affiliation{Department of Energy Conversion and Storage, Technical University of Denmark, 2800 Kgs.\ Lyngby, Denmark}

\author{Damon J. Carrad}
\affiliation{Department of Energy Conversion and Storage, Technical University of Denmark, 2800 Kgs.\ Lyngby, Denmark}

\author{Emiliano Di Gennaro}
\affiliation{Dipartimento di Fisica "Ettore Pancini", Università degli Studi di Napoli “Federico II”,  Complesso Universitario di Monte S. Angelo, Via Cinthia, 80126 Naples, Italy}

\author{Yu Chen}
\affiliation{CNR-SPIN, c/o Complesso di Monte S. Angelo, via Cinthia, 80126  Naples, Italy}

\author{Nicola Manca}
\affiliation{CNR-SPIN, C.so F. M. Perrone 24, 16152 Genova, Italy}

\author{Nicolas Bergeal}
\affiliation{Laboratoire de Physique et d’Etude des Matériaux, ESPCI Paris, Université PSL, CNRS, Sorbonne Université, Paris, France}

\author{Manuel Bibes}
\affiliation{Laboratoire Albert Fert, CNRS, Thales, Université Paris-Saclay, 91767 Palaiseau, France}

\author{Alexei Kalaboukhov}
\affiliation{Department of Microtechnology and Nanoscience (MC2), Chalmers University of Technology, 41296 Gothenburg, Sweden}

\author{Marco Salluzzo}
\affiliation{CNR-SPIN, c/o Complesso di Monte S. Angelo, via Cinthia, 80126  Naples, Italy}

\author{Roberta Citro}
\affiliation{Dipartimento di Fisica “E.R. Caianiello,” Universita di Salerno, 84084 Fisciano, Italy} \affiliation{CNR-SPIN, c/o Universita di Salerno, 84084 Fisciano, Italy}

\author{Felix Trier}
\affiliation{Department of Energy Conversion and Storage, Technical University of Denmark, 2800 Kgs.\ Lyngby, Denmark}

\author{Nini Pryds}
\affiliation{Department of Energy Conversion and Storage, Technical University of Denmark, 2800 Kgs.\ Lyngby, Denmark}

\author{Fabio Miletto Granozio}
\affiliation{CNR-SPIN, c/o Complesso di Monte S. Angelo, via Cinthia, 80126  Naples, Italy}

\author{Alessia Sambri}
\affiliation{CNR-SPIN, c/o Complesso di Monte S. Angelo, via Cinthia, 80126  Naples, Italy}

\author{Thomas S. Jespersen}
\affiliation{Department of Energy Conversion and Storage, Technical University of Denmark, 2800 Kgs.\ Lyngby, Denmark}\affiliation{Center For Quantum Devices, Niels Bohr Institute, University of Copenhagen, 2100 Copenhagen, Denmark} 

%%%%%%%%%%%%% TITLE %%%%%%%%%%%%%%%%%%%%%
\title{Supplementary Information for Intrinsic Negative-U Centers in Freestanding LaAlO$_3$/SrTiO$_3$ Micro-membranes}

\maketitle

\onecolumngrid

%S1
\section{Determining the orientation of LAO/STO membranes relative to the substrate} \label{Determining the orientation of LAO/STO membranes relative to the substrate}

The micro-membranes were grown with the LAO layer on top of the STO layer, resulting in an upward curvature ($\cup$-shaped) induced by the strain due to the smaller lattice constant of LAO \cite{sambri2020self}. During the transfer from the growth substrate to the device substrate, membranes could be flipped, resulting in a downward curvature ($\cap$-shaped) with the LAO layer facing the substrate. To determine the orientation, we examined the curvature using tilted scanning electron microscopy (SEM) and performed energy-dispersive X-ray spectroscopy (EDX) on the top surface. SEM revealed that Dev\#1 exhibited a $\cap$-shaped curvature, indicating the STO layer on top (Fig.\ S1a), while Dev\#2 showed a $\cup$-shaped curvature, with the LAO layer on top (Fig.\ S1b). Correspondingly, EDX spectra (Fig.\ S1c) show a high Sr signal for Dev\#1, whereas Dev\#2 exhibits lower Sr intensity and higher La and Al peaks. These observations confirm that Dev\#1 had the STO layer on top ($\cap$) and Dev\#2 had the LAO layer facing up ($\cup$).

%Figure 
\begin{figure}[H]
\centering
\includegraphics[width = \linewidth]{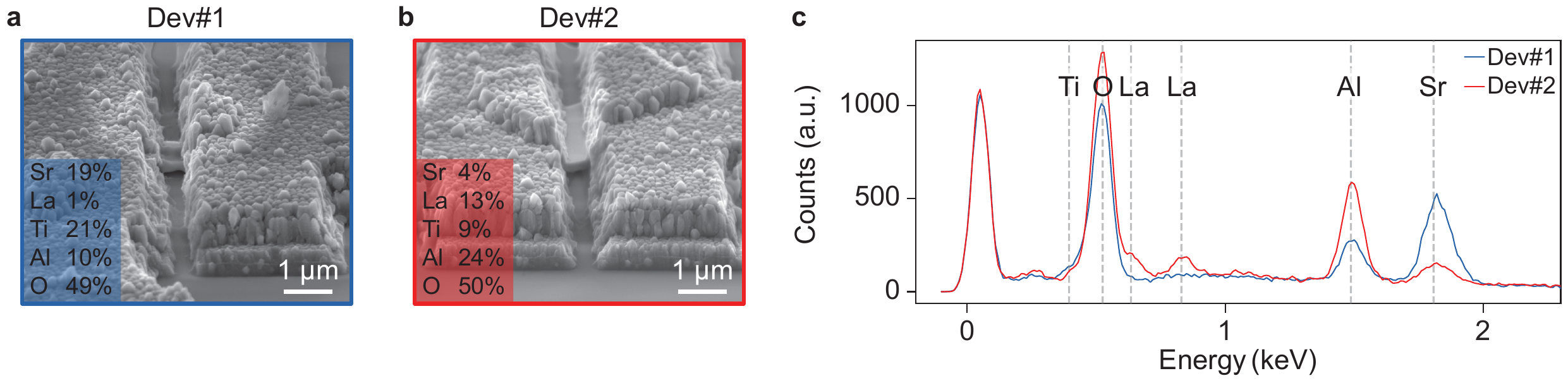}
\caption{
\textbf{a} Titled-SEM of Dev\#1.  
\textbf{b} Titled-SEM of Dev\#2.   
\textbf{c} EDX spectra acquired from the top of Dev\#1 and Dev\#2.  The atomic compositions derived from the spectra are shown in panels a and b.
}

\end{figure} 

%S2
\section{Two-terminal measurements and series resistance} \label{Two-terminal measurements and series resistance}

The two-terminal configuration introduced a series resistance, $R_{\text{s}}$, from the contacts and cryostat filters, which can be measured as the baseline value in the superconducting regime (Fig.\ S2a,c). We estimated $R_{\text{s}}$, from the slope of a linear fit to total measured DC voltage, $V_{\text{DC,tot}}$, versus the applied DC current, $I_{\text{DC}}$, in the superconducting regime (Fig.\ S2a) and subtracted it from the measured resistance (Fig.\ S2d). $R_{\text{s}}$ ranges between 6 and 11 k$\Omega$, with the large device-to-device variation expected from differences in the cryostat filter resistance ($6.07-10.66$ k$\Omega$). \\
\indent In voltage biased measurements the voltage applied drops both on $R_{\text{s}}$ and on the device. To find the effective bias on the device we subtracted $R_{\text{s}}\cdot I_{\text{DC}}$ from the applied value (Fig.\ S2b).

%Figure 
\begin{figure}[H]
\centering
\includegraphics[width = \linewidth]{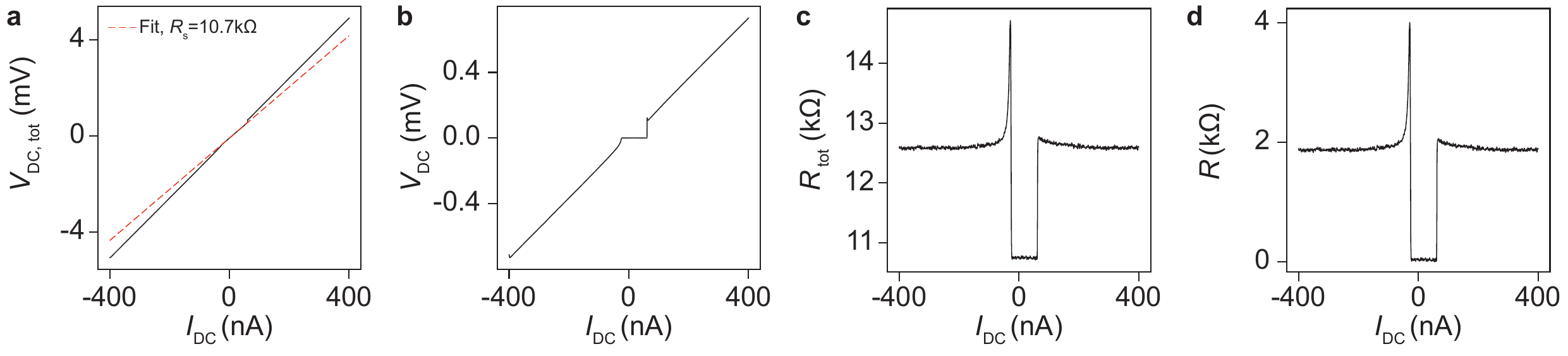}
\caption{
\textbf{a} Total DC voltage drop, $V_{\text{SD,tot}}$, measured as a function of applied bias current, $I_{\text{DC}}$ (black solid line), with a linear fit at low bias used to extract the series resistance $R_{\text{s}}$ (red dashed line).  
\textbf{b} Voltage drop across the device, $V_{\text{SD}} = V_{\text{SD,tot}} - R_{\text{s}} \cdot I_{\text{DC}}$, as a function of $I_{\text{DC}}$.  
\textbf{c} Total differential resistance, $R_{\text{tot}}$, measured as a function of $I_{\text{DC}}$.  
\textbf{d} Differential resistance of the device, $R = R_{\text{tot}} - R_{\text{s}}$, as a function of $I_{\text{DC}}$.  
}

\end{figure}

%\pagebreak

%S3
\section{Calculation of superconducting parameters}
%Figure 
\begin{figure}[H]
\centering
\includegraphics[width = \linewidth]{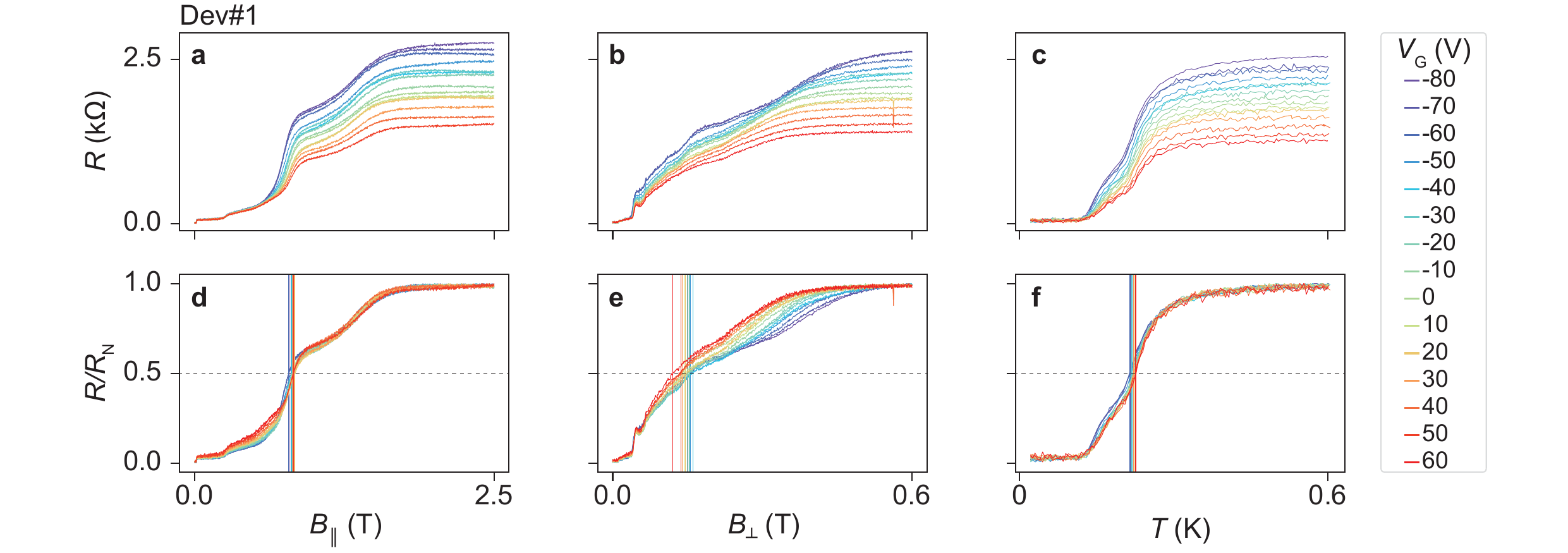}
\caption{ Differential resistance of Dev\#1 as a function of \textbf{a} in-plane magnetic field, $B_{\parallel}$, \textbf{b} out-of-plane magnetic field, $B_{\perp}$, and \textbf{c} temperature, $T$, for selected $V_{\text{G}}$. Panels (\textbf{d–f}) display the corresponding traces normalized to the maximum resistance for each trace. The 50\% threshold  is indicated by a black horizontal line, and the extracted critical values are marked by vertical lines.}
\end{figure} 
This section details how the parameters of the superconducting phase discussed in the main text  were determined:
\begin{itemize}
\item The switching current, $I_{\text{s}}$, was obtained from the data in Fig.\ 1c as the bias current at which the resistance exhibits a maximum when increasing $I_{\text{SD}}$ from the superconducting to the resistive state. 
\item The normal-state resistance, $R_{\text{N}}$, was defined as the zero-bias resistance at $B_{\perp} = 1$T, where superconductivity was fully suppressed.
\item  The critical in-plane ($B_{\parallel}^\text{c}$) and out-of-plane ($B_{\perp}^\text{c}$) magnetic fields were extracted from the zero-bias resistance traces versus $B$ (Fig.\ S3a-b). The critical field was defined as the value of $B$ at which the resistance reached half of its normal-state value, taken as the resistance measured at the maximum applied field, $B_{\parallel} = 2.5$ T, $B_{\perp} = 0.6$ T (Fig.\ S3d,e). The $V_\text{G}$-dependence of $B_{\parallel}^\text{c}$ and $B_{\perp}^\text{c}$ is shown in Fig.\ S4.
\item The superconducting coherence length was estimated as $\xi = \sqrt{\Phi_0/(2\pi B_{\perp}^{c})}$, and the effective thickness of the superconducting layer as $d = \sqrt{3} \, \Phi_0/(\pi \, \xi \, B_{\parallel}^\text{c})$, where $\Phi_0$ is the magnetic flux quantum and $\mu_0$ is the vacuum permeability \cite{reyren2009anisotropy}. We note that, because the freestanding membranes are curved, the $x$-axis may not have been perfectly aligned with the 2DEG plane, potentially leading to an underestimation of $B_{\parallel}^\text{c}$. Accordingly, the extracted thickness $d$ should be regarded as an upper bound. 
\item The critical temperature, $T_{\text{c}}$, was extracted from the zero-bias resistance traces versus $T$ at  $B_{\parallel}=25$ mT (Fig.\ S3c), as the value of $T$ at which the resistance reached half of its normal-state value, taken as the resistance measured at $T = 600$ mK (Fig.\ S3f). The $V_\text{G}$-dependence of $T_\text{c}$ is shown in Fig.\ S4.
\end{itemize}

\begin{figure}[H]
\centering
\includegraphics[width = \linewidth]{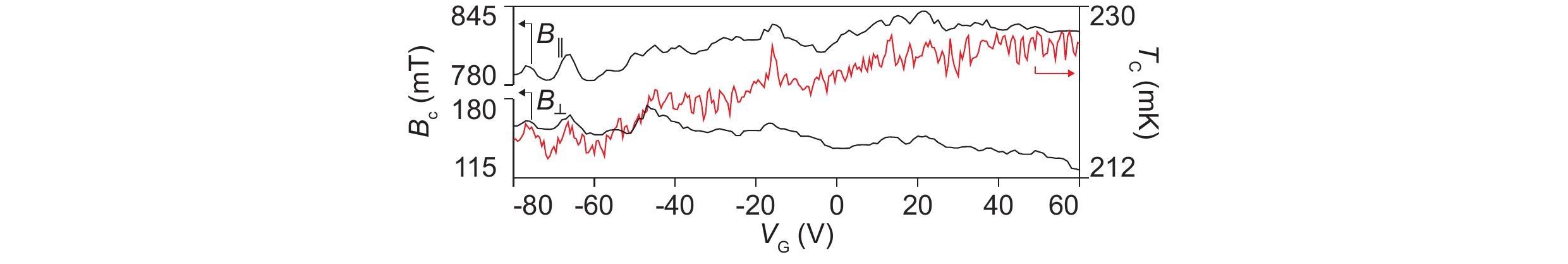}
\caption{Critical magnetic field and critical temperature as a function of $V_\text{G}$. }

\end{figure}

%S4
\section{Background conductance fitting procedure}

Unlike the behavior expected for an isolated quantum dot, the measured conductance does not vanish between the conductance peaks. This is attributed to a parallel conduction channel, which provides a gate-dependent background conductance associated with increasing puddle coupling. To isolate the contribution of center, this background is subtracted from the data. For each voltage bias or magnetic field, the conductance trace as a function of gate voltage is fitted with a fifth-order polynomial, and the resulting background is removed from the corresponding line cut. An example of such a fit is shown in Fig.\ S5.
\begin{figure}[H]
\centering
\includegraphics[width =\linewidth]{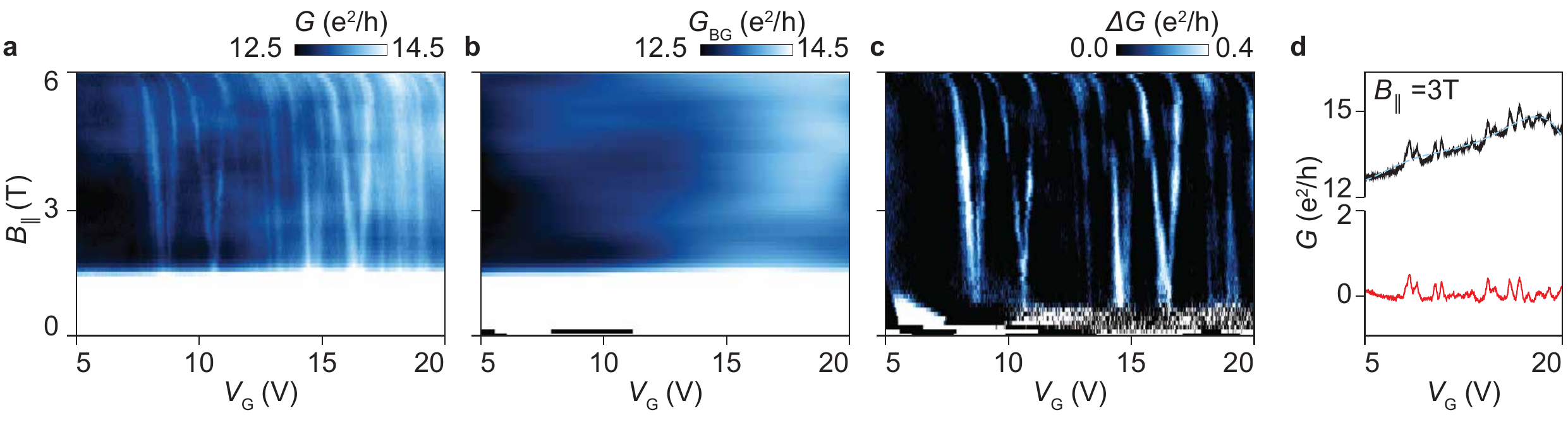}
\caption{
\textbf{a} Total zero-bias differential conductance of Dev\#1 as a function of $V_\text{G}$ and $B_{\parallel}$.
\textbf{b} Background differential conductance extracted from fifth-order polynomial fits to each trace, at fixed magnetic field.
\textbf{c} Residual differential conductance after subtracting the background.
\textbf{d} Representative line cuts at $B_{\parallel}=3$ T showing the measured conductance (black), the polynomial background (blue), and the residuals (red).
}
\end{figure}

%S5
\section{Negative-$U$ center in Dev\#2}

Figure S6a shows the residual conductance of Dev\#2 as a function of $V_\text{G}$ and $V_\text{SD}$, at $B_{\perp} = 0.1$ T. Similarly to what presented in the main text for Dev\#1, the measurements show diamonds of enhanced conductance, indicative of quantum dot behavior. Figures S6b and S6c display the evolution of the zero-bias conductance peaks with magnetic field, again consistent with the behavior observed for Dev\#1. The characteristics energies of the device are discussed in the main text and analyzed in Fig.\  3.\\
\indent Figure S7 presents additional measurements of the differential conductance of Dev\#2 as a function of $V_\text{G}$ and $V_\text{SD}$  at different magnetic fields, further illustrating the progressive splitting of the conductance diamonds with increasing field.\\
\indent Figure S8 shows the behavior of Dev\#2 after thermal cycling. In contrast to Dev\#1, the conductance diamonds disappear after cycling to 125 K.
%Figure 
\begin{figure}[H]
\centering
\includegraphics[width = \linewidth]{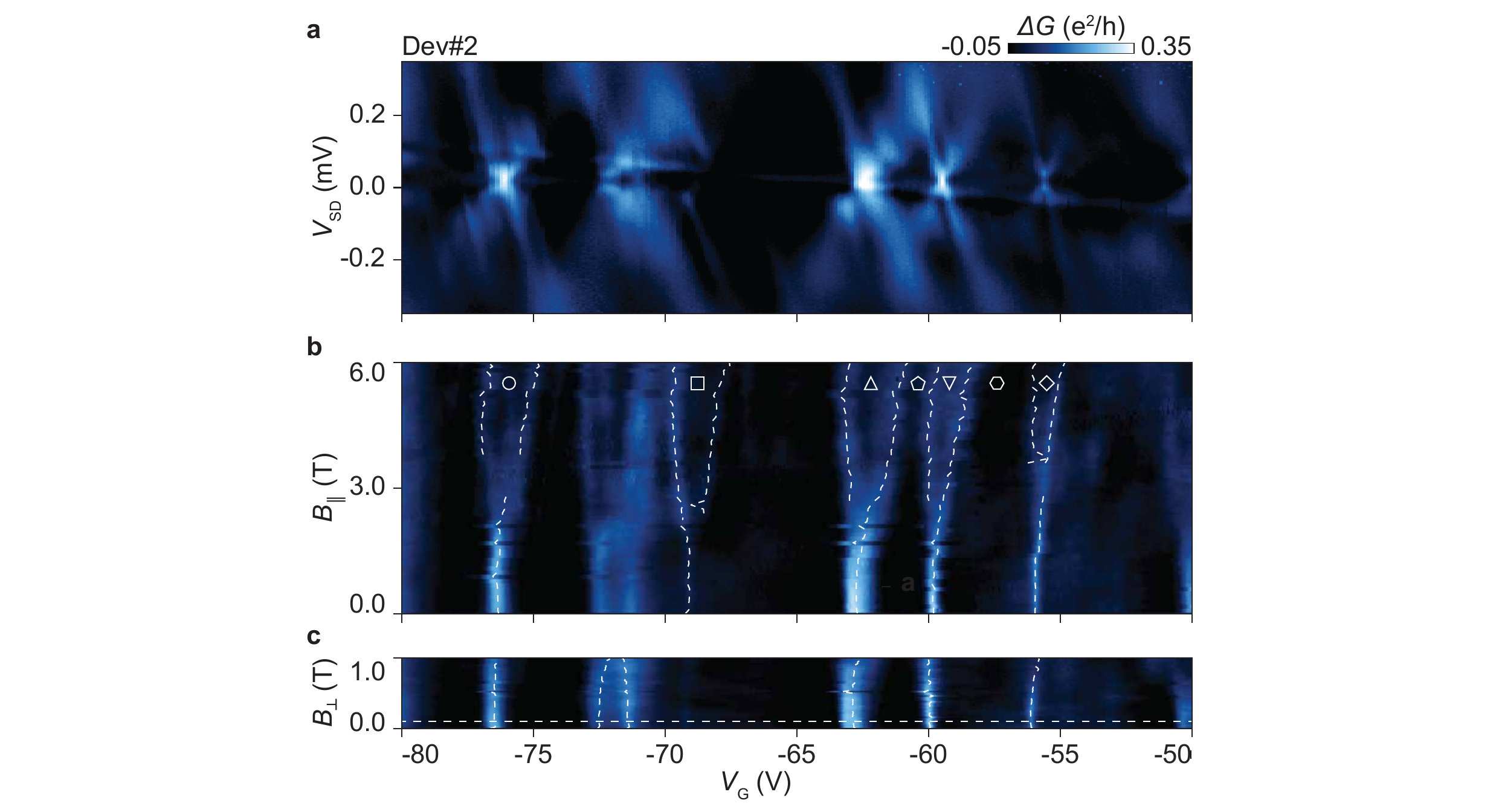}
\caption{
\textbf{a} Residual conductance, $\Delta G$, of Dev\#2 as a function of $V_\text{G}$ and $V_\text{SD}$, at $B_{\perp} = 0.1$ T (horizontal white dashed line in panel c). 
\textbf{b} Zero-bias $\Delta V_\text{G}$ versus $V_\text{G}$ and $B_{\parallel}$.   The white dashed lines show the conductance peak positions obtained from single or double Lorentzian fits.  The symbols correspond to those in Fig.\ 3.
\textbf{c}  Zero-bias $\Delta V_\text{G}$ versus $V_\text{G}$ and $B_{\perp}$.   The white dashed lines show the conductance peak positions obtained from single Lorentzian fits. }
\end{figure}
\begin{figure}[H]
\centering
\includegraphics[width = \linewidth]{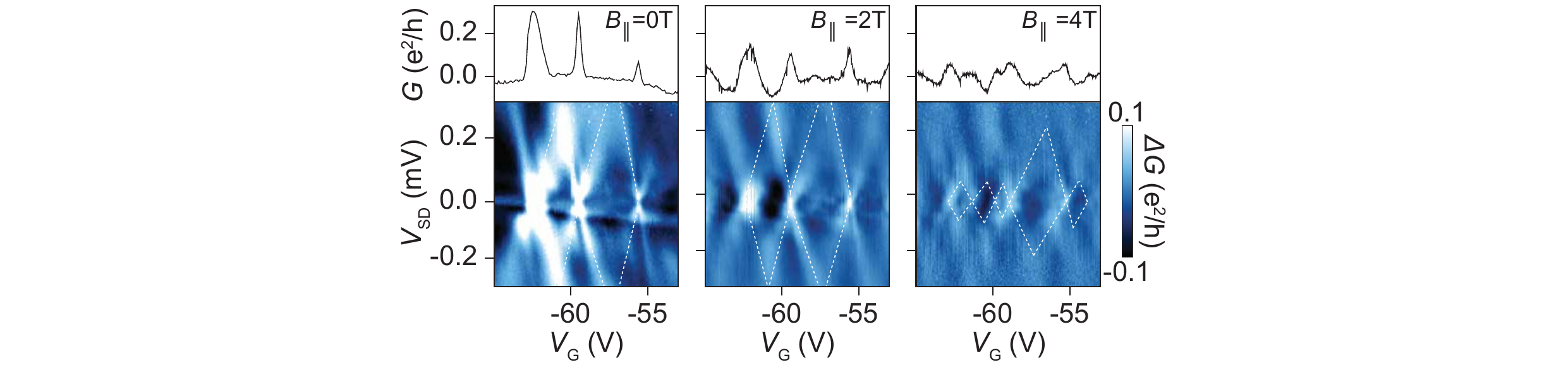}
\caption{Residual conductance, $\Delta G$, of Dev\#2 versus $V_\text{G}$ and $V_\text{SD}$, at $B_{\parallel} = 0$  ($B_{\perp} = 0.1$ T),  $B_{\parallel} = 2$ T and $B_{\parallel} = 4$ T.  White dashed lines are guides to the eye highlighting the conductance diamonds. The top insets show the zero-bias traces. }
\end{figure}

\begin{figure}[H]
\centering
    \includegraphics[width = \linewidth]{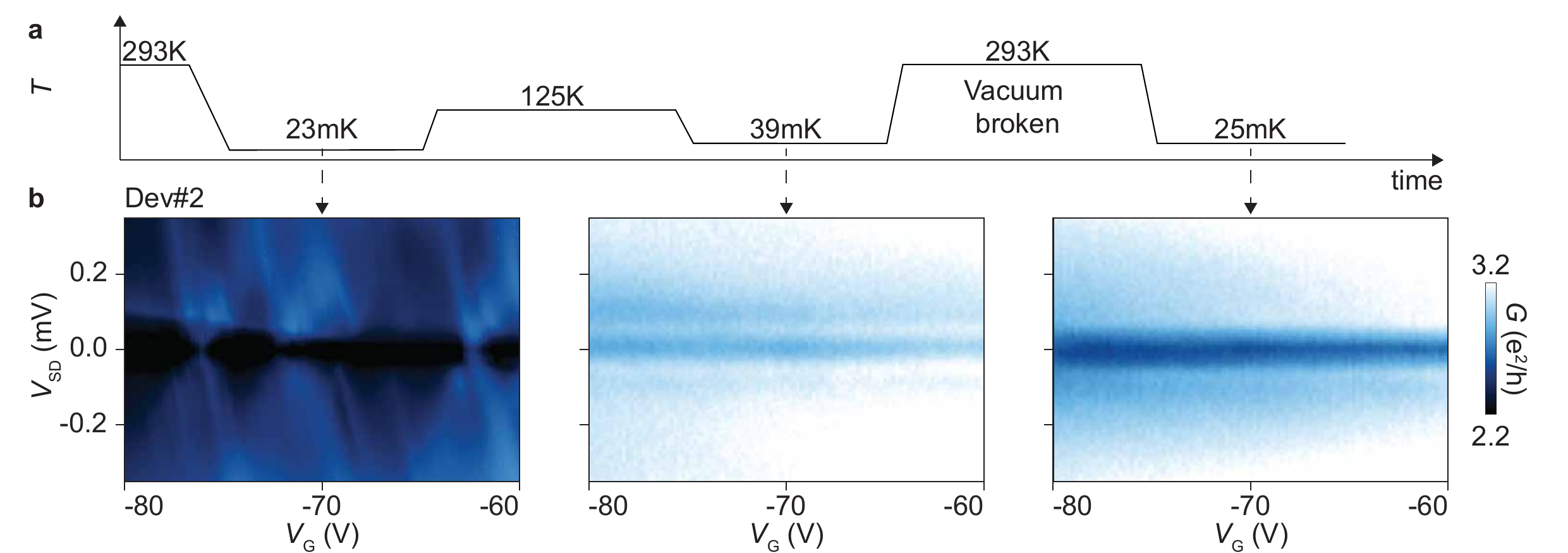}
    \caption{\textbf{a} Thermal cycling sequence applied to the devices.
    \textbf{b} Total differential conductance of Dev\#2 after the first cool-down, after thermal cycling to 125 K, and after thermal cycling to room temperature with exposure to air,  at $B_{\perp} = 0.1$ T.  }

\end{figure}

%S6
\section{Peak fitting procedure}
The zero-bias conductance peaks shown in Fig.\ 2 and Fig.\ S6 were analyzed using a two-step fitting procedure. Peak positions were first estimated for each magnetic field using a peak-finding algorithm, providing initial values for subsequent Lorentzian fits. When two peaks appeared in close proximity, i.e. after peak splitting, a double Lorentzian fit was applied; otherwise, a single Lorentzian was used. An example of the fitting procedure for Dev\#1 is presented in Fig.\ S9.
%Figure 
\begin{figure}[H]
\centering
\includegraphics[width = \linewidth]{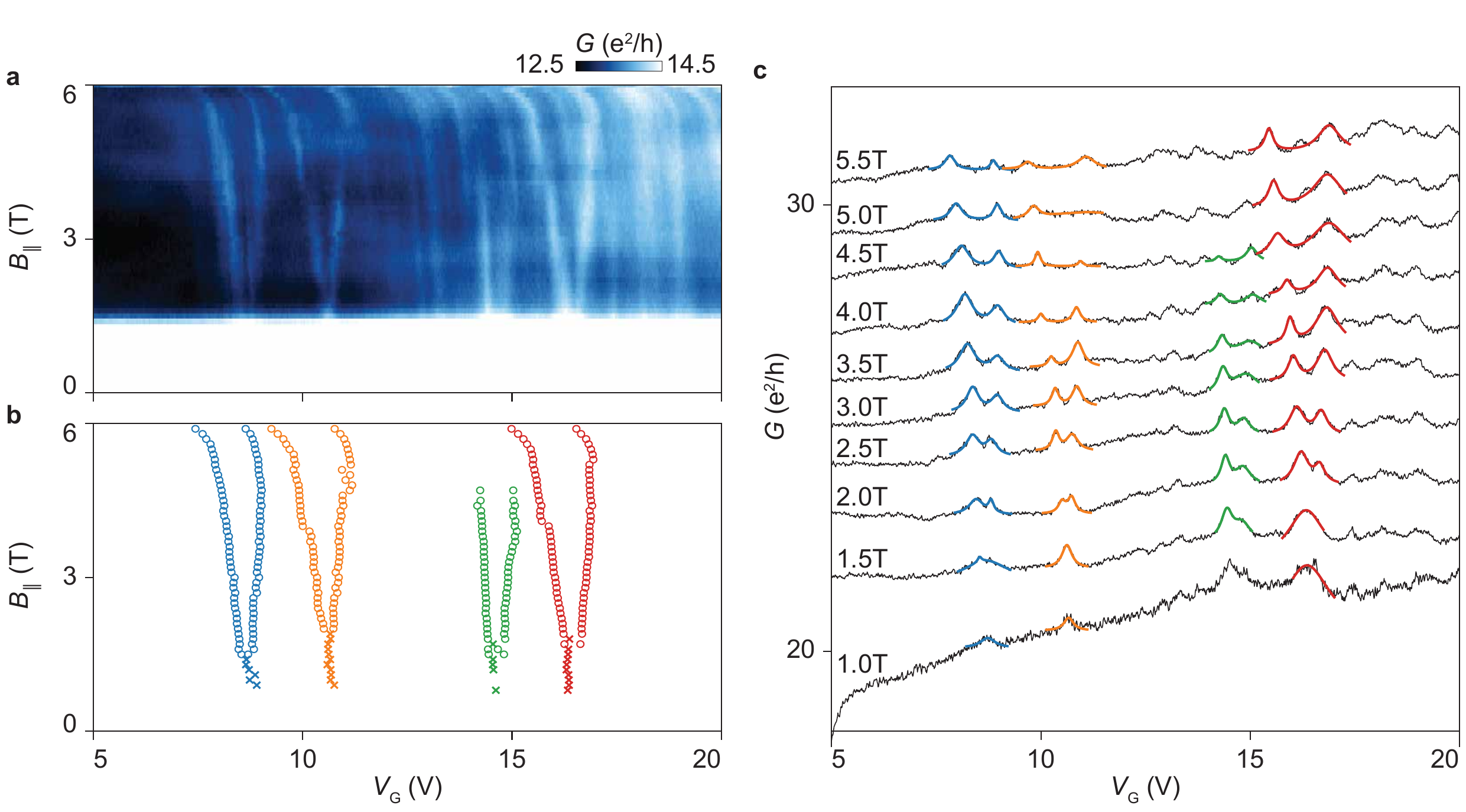}
\caption{
\textbf{a} Zero-bias total differential conductance of Dev\#1 as a function of $V_\text{G}$ and $B_{\parallel}$.  
\textbf{b} Peak positions obtained from Lorentzian fits: crosses correspond to single-Lorentzian fits, circles to double-Lorentzian fits.  
\textbf{c} Conductance traces from panel a at selected $B_{\parallel}$, together with the corresponding single or double Lorentzian fits. The traces have been vertically offset for clarity.
}
\end{figure}

%S7
\section{Transport simulation of a negative-$U$ center}
The transport through the negative-$U$ center is modeled with a two-orbital Anderson model using the Qmec package \cite{KIRSANSKAS2017317}. The Hamiltonian of the quantum dot coupled to the leads is defined by:
\begin{align}
H = H_{QD} + H_{L} + H_{t}.
\end{align}
Here the Hamiltonian of the isolated quantum dot is:
\begin{align}
H_{QD} &= \sum_{i,\sigma} \varepsilon_i\, d_{i\sigma}^\dagger d_{i\sigma}
\;+\;
\frac{1}{2}\sum_i g_i \mu_B B 
\left( d_{i\uparrow}^\dagger d_{i\uparrow} - d_{i\downarrow}^\dagger d_{i\downarrow} \right) + \sum_i U_i\, n_{i\uparrow} n_{i\downarrow}
\;+\;
\sum_{i \neq j} \hat{n}_i \hat{n}_j\, V_{ij} \, ,
\end{align}
where $\varepsilon_i$ is the energy of the $i$-th orbital given by: $\varepsilon_i = \alpha V_\text{G} + \alpha V_0$, where $\alpha$ is the lever arm, $V_\text{G}$ is the gate voltage and $V_0$ is the voltage offset. \\
The operator $d_{i\sigma}^\dagger$ creates a fermion on the quantum dot with spin $\sigma$, so the first term describes the energy cost of adding an electron to the center. The second term describes the Zeeman energy splitting for orbital $i$ under a magnetic field $B$. The third term represents the on-site Hubbard interaction for orbital $i$. The last term accounts for the inter-orbital charging energy between different orbitals $i \neq j$. \\
The Hamiltonian of the leads is :
\begin{align}
H_{L} &= \sum_{\alpha,k} \varepsilon_k\, c_{\alpha,k}^\dagger c_{\alpha,k} \, ,
\end{align}
where the operator $c_{\alpha,k}^\dagger$ creates an electron in lead $\alpha$ with momentum $k$ and $\varepsilon_k$ is the corresponding energy. \\
Each lead is coupled to the quantum dot through the following tunneling Hamiltonian:
\begin{align}
H_{t} &= \sum_{i,\alpha,k} t_{\alpha,k,i}\, c_{\alpha,k}^\dagger d_{i\sigma} + \text{H.c.} \, ,
\end{align}
where $t_{\alpha,k,i, \sigma}$ is the coupling between state $k, \alpha$ in the lead and state $i, \sigma$ in the quantum dot, which is related to the tunneling rate $\Gamma_\alpha =2\pi \lvert t_{\alpha,k,i, \sigma}\rvert^2$. \\
We calculate the conductance through the dot as function of $V_\text{G}$, $B$ and voltage bias $V_\text{SD}$ using the Pauli master equation approach, with the corresponding assumptions described in Ref. \citenum{KIRSANSKAS2017317}. Within this approach, the states in the leads are occupied according to the Fermi-Dirac distribution: $f_\alpha(\varepsilon) = \left(e^{(\varepsilon - \mu_\alpha)/k_bT_\alpha}\right)^{-1} $, where $\mu_\alpha$ and $T_\alpha$ are the chemical potential and temperature in lead $\alpha$. To incorporate the voltage bias, we symmetrically change the chemical potential around zero in both leads.\\
Table \ref{tbl:params_model} presents the parameters of the model that represent the experimental negative-$U$ center as described in the main text. For the calculation, the parameters are scaled for numerical stability.

\begin{table}[H]
\label{tbl:params_model}
\begin{center}

\begin{tabular}{|l|l|}
\hline
\textbf{Parameter} & \textbf{Value}  \\ \hline
    $\alpha$               &    \SI{10.24e-5}{\eV\per\volt}           \\ \hline
    $V_0$               &   \SI{12}{\volt}             \\ \hline
    $g_1$               &     \num{0.36}          \\ \hline
     $g_2$               &    \num{0.59}               \\ \hline

     $U_1$               &      \SI{-4.49e-6}{\eV}          \\ \hline
     $U_2$               &        \SI{-50.28e-6}{\eV}         \\ \hline
     $V_{12}$               &       \SI{210e-6}{\eV}          \\ \hline
     $V_{21}$               &      \SI{210e-6}{\eV}           \\ \hline
       $t_{\alpha,k,i}$      &        \SI{3.592e-6}{\eV}        \\ \hline

     $\Gamma_{L,R}$               &      \SI{4.05e-11}{\eV}           \\ \hline
    $T_{L,R}$               &      \SI{0.1}{\kelvin}           \\ \hline

\end{tabular}
\end{center}
\caption{Parameters used for the transport simulations presented in Fig.\ 2d-f.}
\end{table}

%S8
\section{Electrostatic simulation of the center size}

To estimate the size of the center, electrostatic simulations are performed as described in the main text. The results reported in the main text are obtained for a gate voltage $V_\text{G} = 10\ \mathrm{V}$ and a micromembrane with the STO layer facing the substrate. 
However, the results can be generalized to devices with the LAO layer facing the substrate and to a wider gate voltage range spanning the experimental conditions of both Dev\#1 and  Dev\#2.  This is because the thick SiO$_2$ dielectric dominates the electrostatics.  Figure S10 shows the simulated capacitance of  70 nm layers of LAO and STO as a function of $V_\text{G}$. 
While the intrinsic capacitances of LAO and STO differ by orders of magnitude and the STO capacitance is gate-dependent, introducing the 500 nm SiO$_2$ layer between the oxide layer and the gate makes the overall capacitances of the two configurations nearly identical and independent of $V_\text{G}$ over the range 1 $\mu$V -- 100 V. 
This indicates that the choice of $V_\text{G}$ and the orientation of the micromembrane have negligible effect on the simulations.  

To further confirm this, the simulations shown in Fig.\ 5a were repeated with the opposite micromembrane orientation, and the results are presented in Fig.\ S11.

%Figure 
\begin{figure}[H]
\centering
\includegraphics[width = \linewidth]{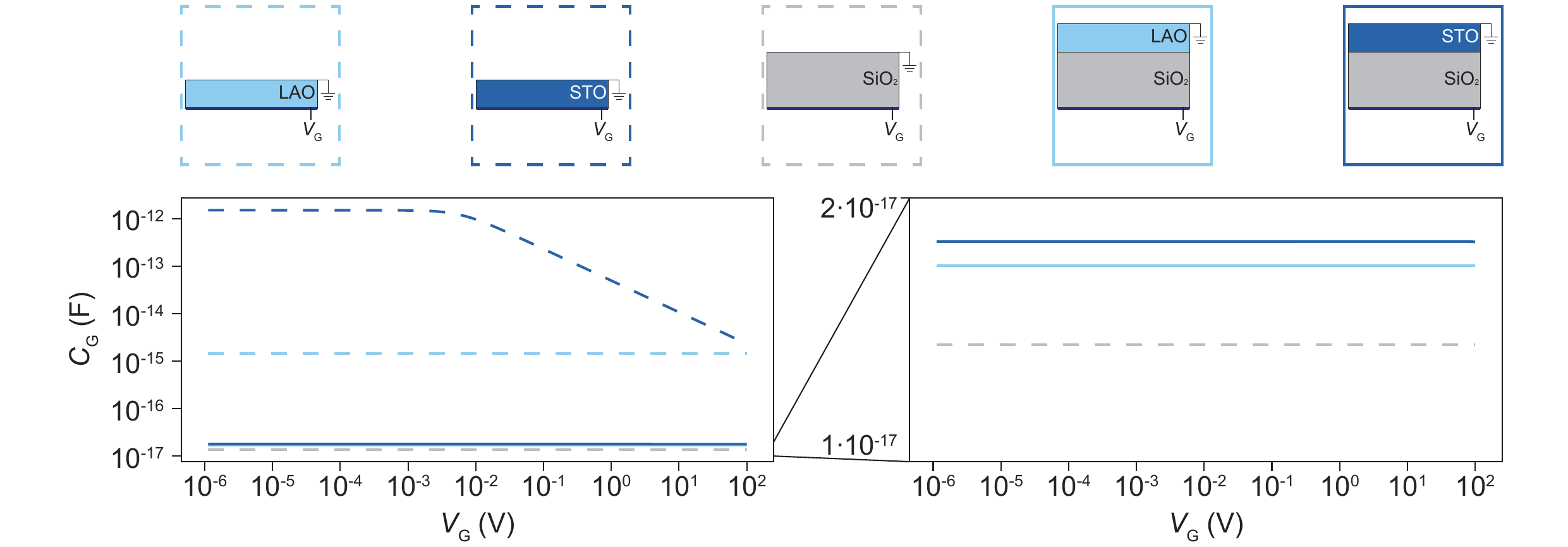}
\caption{ Simulated capacitance as a function of $V_\text{G}$ for single layers of LAO (70 nm), STO (70 nm), and SiO$_2$ (500 nm), shown with dashed lines, and for bilayers of LAO (70 nm)/SiO$_2$ (500 nm) and STO (70 nm)/SiO$_2$ (500 nm), shown with solid lines. The schematics on top illustrate the side view of the geometries used in the simulations.
}
\end{figure}

%Figure 
\begin{figure}[H]
\centering
\includegraphics[width = \linewidth]{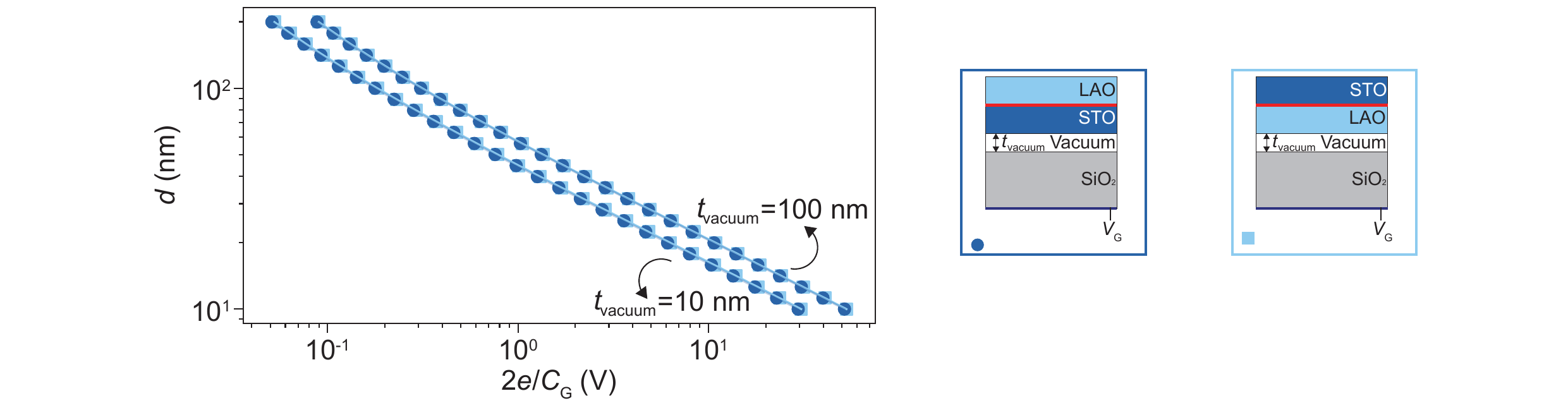}
\caption{ Estimated center diameter as a function of gate capacitance $C_\text{G}$ for a contact aspect ratio $w/w_\text{t} = 0.2$ and vacuum thicknesses $t_{\text{vacuum}}=10$ and 100 nm, for two micromembrane orientations: STO on the substrate (circles) and LAO on the substrate (squares).
}
\end{figure}

\section{Surface roughness}
Figure S12a and S12b show atomic-force micrographs of two micromembranes on Si, one with the LAO side facing up and one with the STO side facing up. These images illustrate the typical membrane curvature discussed in Section S1, as well as the characteristic surface roughness. As expected from the spalling-based growth, the STO surface appears rougher than the LAO surface. Figure S12c presents a higher-magnification view of panel b, revealing surface steps and roughness on the order of 5–20 nm.

To assess the influence of this roughness on the 2DEG, Figure S13 shows electrostatic simulations using a surface profile modeled as kinks with a height of 5 nm and a width of 50 nm. The resulting variations of the dielectric constant of STO and the electric displacement field under a gate voltage of 100 V are displayed. The effect of the roughness is minimal and negligible at the height of the 2DEG (70 nm), indicating that this cannot account for the presence of confinement centers.

\begin{figure}[H]
\centering
\includegraphics[width = \linewidth]{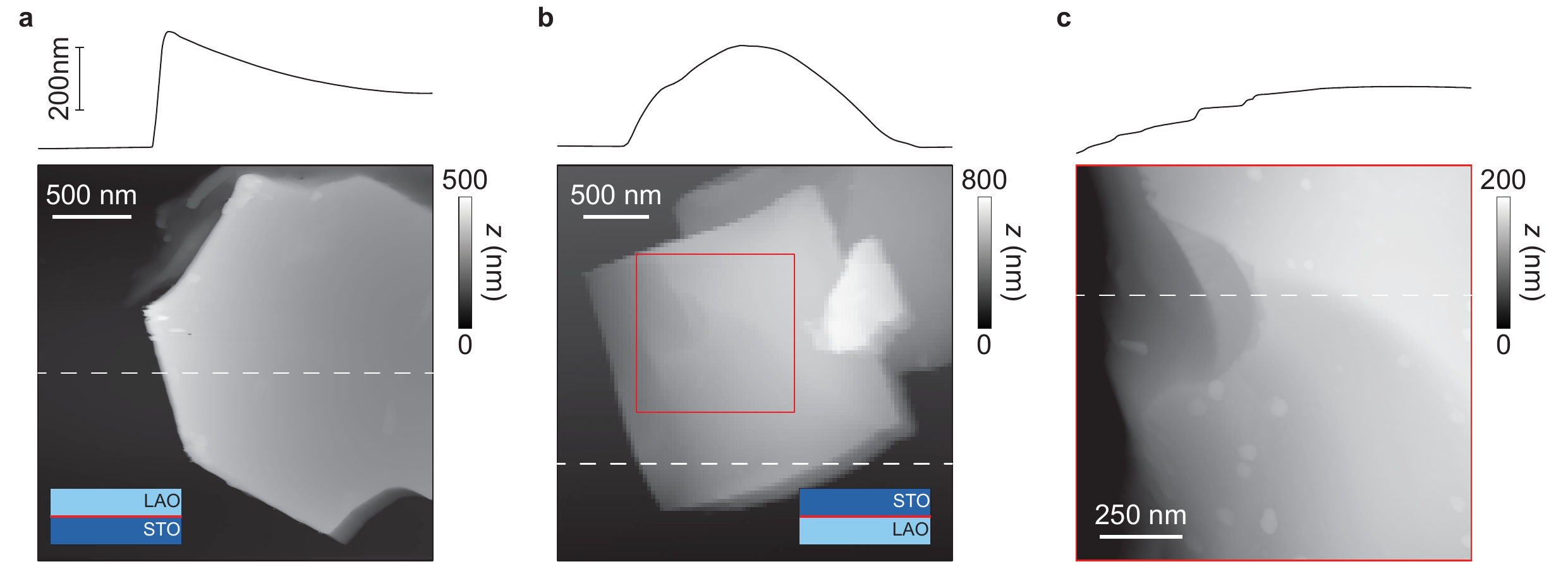}
\caption{ Atomic-force micrograph of LAO/STO micromembrane on Si. \textbf{a} Membrane with the LAO side facing up. \textbf{b} Membrane with the STO side facing up. \textbf{c} Zoom-in of panel b. The inset at the top shows height profiles taken along the white dashed lines in each panel. 
}
\end{figure}

\begin{figure}[H]
\centering
\includegraphics[width = \linewidth]{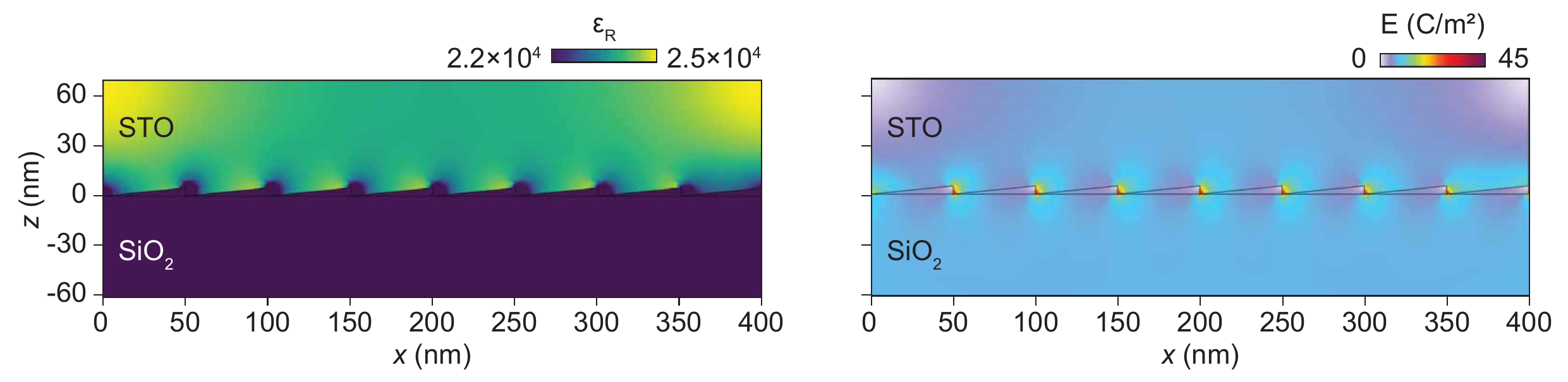}
\caption{ \textbf{a} Dielectric constant of STO and \textbf{b} electric displacement field, calculated under an applied gate voltage of 100 V including a surface roughness modeled as kinks with a height of 5 nm and a width of 50 nm, as indicated by the black profile in each panel.
}
\end{figure}

\bibliography{ref}